\documentclass[iop]{emulateapj}
\usepackage{amssymb,amsmath,amsthm}
\usepackage{natbib}
\usepackage{xcolor}
\usepackage{graphicx}

\newcommand\ie{i.e.}

\newcommand\kms{\ifmmode{\rm km\ s^{-1}}\else$\rm km\ s^{-1}$\fi}
\newcommand\ergs{\ifmmode{\rm erg\ s^{-1}}\else$\rm erg\ s^{-1}$\fi}
\newcommand\mydeg{\ifmmode{^\circ}\else$^\circ$\fi}
\def\eps@scaling{1.0}
\newcommand\plotthree[3]{{
 \typeout{Plotthree included the files #1 #2 #3}
 \centering
 \leavevmode
 \columnwidth=.33\columnwidth
 \includegraphics[width={\eps@scaling\columnwidth}]{#1}
 \includegraphics[width={\eps@scaling\columnwidth}]{#2}
 \includegraphics[width={\eps@scaling\columnwidth}]{#3}
}}
\newcommand\plotrtwo[2]{{
 \typeout{Plotrtwo included the files #1 #2}
 \centering
 \leavevmode
 \columnwidth=.5\columnwidth
 \includegraphics[angle=270, width={\eps@scaling\columnwidth}]{#1}
 \includegraphics[angle=270, width={\eps@scaling\columnwidth}]{#2}
}}
\newcommand\plotfour[4]{{
 \typeout{Plotfour included the files #1 #2 #3 #4}
 \centering
 \leavevmode
 \columnwidth=.5\columnwidth
 \includegraphics[width={\eps@scaling\columnwidth}]{#1}
 \includegraphics[width={\eps@scaling\columnwidth}]{#2}
 \includegraphics[width={\eps@scaling\columnwidth}]{#3}
 \includegraphics[width={\eps@scaling\columnwidth}]{#4}
}}

\shorttitle{A Dissociative Merger in CIZA J0107}
\shortauthors{RANDALL ET AL.}
\slugcomment{Accepted version of \today}

\begin{document}

\title{Multi-wavelength Observations of the Dissociative Merger in the
  Galaxy Cluster CIZA~J0107.7+5408}

\author{S. W. Randall\altaffilmark{1},
  T. E. Clarke\altaffilmark{2},
  R. J. van Weeren\altaffilmark{1},
  H. T. Intema\altaffilmark{3},
  W. A. Dawson\altaffilmark{4},
  T. Mroczkowski\altaffilmark{5},
  E. L. Blanton\altaffilmark{6},
  E. Bulbul\altaffilmark{7},
  S. Giacintucci\altaffilmark{8}}

\altaffiltext{1}{Harvard-Smithsonian Center for Astrophysics, 60
  Garden St., Cambridge, MA 02138, USA; srandall@cfa.harvard.edu}
\altaffiltext{2}{Naval Research Laboratory, 4555 Overlook Avenue SW,
  Code 7213, Washington, DC 20375, USA}
\altaffiltext{3}{National Radio Astronomy Observatory, 1003 Lopezville
  Road, Socorro, NM 87801-0387, USA}
\altaffiltext{4}{Lawrence Livermore National Lab, 7000 East Avenue,
  Livermore, CA 94550, USA}
\altaffiltext{5}{U.S. Naval Research Laboratory, 4555 Overlook Avenue SW, Washington, DC 20375, USA}
\altaffiltext{6}{Astronomy Department and Institute for Astrophysical
  Research, Boston University, 725 Commonwealth Avenue, Boston, MA
  02215, USA}
\altaffiltext{7}{Kavli Institute for Astrophysics \& Space Research,
  Massachusetts Institute of Technology, 77 Massachusetts Ave,
  Cambridge, MA 02139, USA}
\altaffiltext{8}{Department of Astronomy, University of Maryland,
  College Park, MD 20742-2421, USA}

\begin{abstract}
We present results based on X-ray, optical, and radio
observations of the massive galaxy cluster CIZA~J0107.7+5408.  We find
that this system is a post core passage, dissociative, binary merger,
with the optical galaxy density peaks of each subcluster leading their
associated X-ray emission peaks.  This separation occurs because the
diffuse gas experiences ram pressure forces while the effectively collisionless
galaxies (and presumably their associated dark matter halos) do not.
This system contains double peaked diffuse radio emission,
possibly a 
double radio relic with the relics lying 
along the merger axis and also leading the X-ray cores.  We find
evidence for a
temperature peak associated with the SW relic, likely created by the
same merger shock that is powering the relic radio emission in this region.
Thus, this system is a relatively rare clean example of a dissociative
binary merger, which can in principle be used to place constraints on the
self-interaction cross-section of dark matter.
Low frequency radio observations reveal ultra-steep spectrum diffuse
radio emission that is not correlated with the X-ray, optical, or high
frequency radio emission.  We suggest that these sources are radio
phoenixes, which are preexisting non-thermal particle
populations that have been re-energized through adiabatic compression
by the same merger shocks that power the radio relics.
Finally, we place upper limits on inverse Compton emission from the SW
radio relic.
\end{abstract}
\keywords{dark matter --- galaxies: clusters: general --- galaxies: clusters:
  individual (CIZA~J0107.7+5408) --- galaxies: clusters: intracluster
  medium --- X-rays: galaxies: clusters}

\section{Introduction} \label{sec:intro}

According to the standard model of hierarchical structure formation,
galaxy clusters largely grow through cluster mergers.  Major mergers
are the most energetic events in the Universe since the Big Bang, and
can release $\ga 10^{64}$~erg in gravitational binding energy.
Such mergers significantly disturb the intracluster medium (ICM) of
galaxy clusters, and are useful for the study of ICM plasma physics,
cluster scaling relations, non-thermal particle populations, the
growth of large scale structure, and the physics of dark matter.

Dissociative mergers occur when the ICM of a merging subcluster
experiences sufficient ram pressure forces to displace it from
the gravitational potential minimum of its dark matter halo.  Such offsets are
observed as separations between the X-ray surface brightness peaks
and/or the optical galaxy density and gravitational lensing mass
peaks, where the
effectively collisionless galaxies are expected to trace their dark
matter halos. Such systems, e.g., the well-known Bullet Cluster
\citep{2002ApJ...567L..27M}, have been used to
demonstrate the existence of dark matter, and to place limits on the
self-interaction cross-section of dark matter, since self-interacting
dark matter would experience ram-pressure-like forces during a merger,
potentially leading to an offset between the optical galaxies and the
dark matter peak
\citep[e.g.,][]{2004ApJ...606..819M,2006ApJ...648L.109C,2006ApJ...652..937B,2007NuPhS.173...28C,2008ApJ...679.1173R,2012ApJ...747L..42D,2015Sci...347.1462H}. 

Merging or otherwise dynamically disturbed systems often host diffuse
radio sources.
These sources can be broadly divided into at least three categories: radio
relics, radio halos, and radio phoenixes \citep[for a review
see][]{2012A&ARv..20...54F}.  Relics are typically found
in the outskirts of clusters, and often have large sizes ($\ga 1$~Mpc)
and polarized emission.  In contrast, halos are centered on cluster
cores, and are typically not strongly polarized. 
The location and morphology of radio phoenixes varies widely, although they
are typically smaller than relics and halos.
Relics, halos, and phoenixes
are thought to be powered by cluster mergers, through direct shock
acceleration, merger induced turbulence, and adiabatic compression by shocks,
respectively
\citep{1987PhR...154....1B,2001MNRAS.320..365B,2001A&A...366...26E,2015ASSL..407..557B}.
Observationally, ultra-steep spectrum (USS) diffuse radio sources with
radio spectral indices $\alpha \la -1.5$\footnote{Where the radio flux
$F_\nu \propto \nu^\alpha$.}
\citep[such as those described by][]{2001AJ....122.1172S}
represent a particular subset of sources.  It is currently unknown
whether these sources are typically weaker ``classical'' radio relics
associated with more minor mergers, radio phoenixes, ``AGN relics''
\citep[associations of old radio plasma left over from a previous AGN
outburst,][]{2012A&A...548A..75M}, or some mix of these three.
New low-frequency radio survey instruments such as LOFAR
\citep{2013A&A...556A...2V} are expected 
to find large numbers of such diffuse USS sources \citep{2011JApA...32..557R}.

Here we present results from {\it Chandra,} X-ray; VLA, WSRT, and GMRT
radio; and INT optical observations of the galaxy cluster CIZA
J0107.7+5408 (hereafter CIZA~0107), at a redshift of $z=0.1066$
~\citep{2002ApJ...580..774E}, which contains both high and low
frequency diffuse radio emission.  
The primary aims of this paper are to characterize the dynamical
state of CIZA~0107; search for merger signatures (such as disturbed
morphology and shock fronts); compare X-ray, optical, and
radio observations to investigate the nature of the diffuse radio
emission; and to evaluate the potential for this system to probe ICM
and dark matter physics.
When corrected to the reference frame
defined by the CMB\footnote{The NASA/IPAC Extragalactic Database (NED)
  is operated by the Jet Propulsion Laboratory, California Institute
  of Technology, under contract with the National Aeronautics and
  Space Administration.}, the redshift corresponds to a luminosity distance of
$D_{\rm L} = 470$~Mpc and a scale of 1.86~kpc~arcsec$^{-1}$ for a cosmology
with  $\Omega_0 = 0.3$, $\Omega_{\Lambda} = 0.7$, and $H_0 =
73$~\kms~Mpc$^{-1}$.
All uncertainty ranges are 68\% confidence intervals (\ie, 1$\sigma$), unless
otherwise stated.

\section{Data Analysis} \label{sec:data}

\subsection{Chandra X-ray Observations} \label{sec:chandra}

\begin{figure}
\hspace{-0.3in}
\includegraphics[width=1.1\columnwidth]{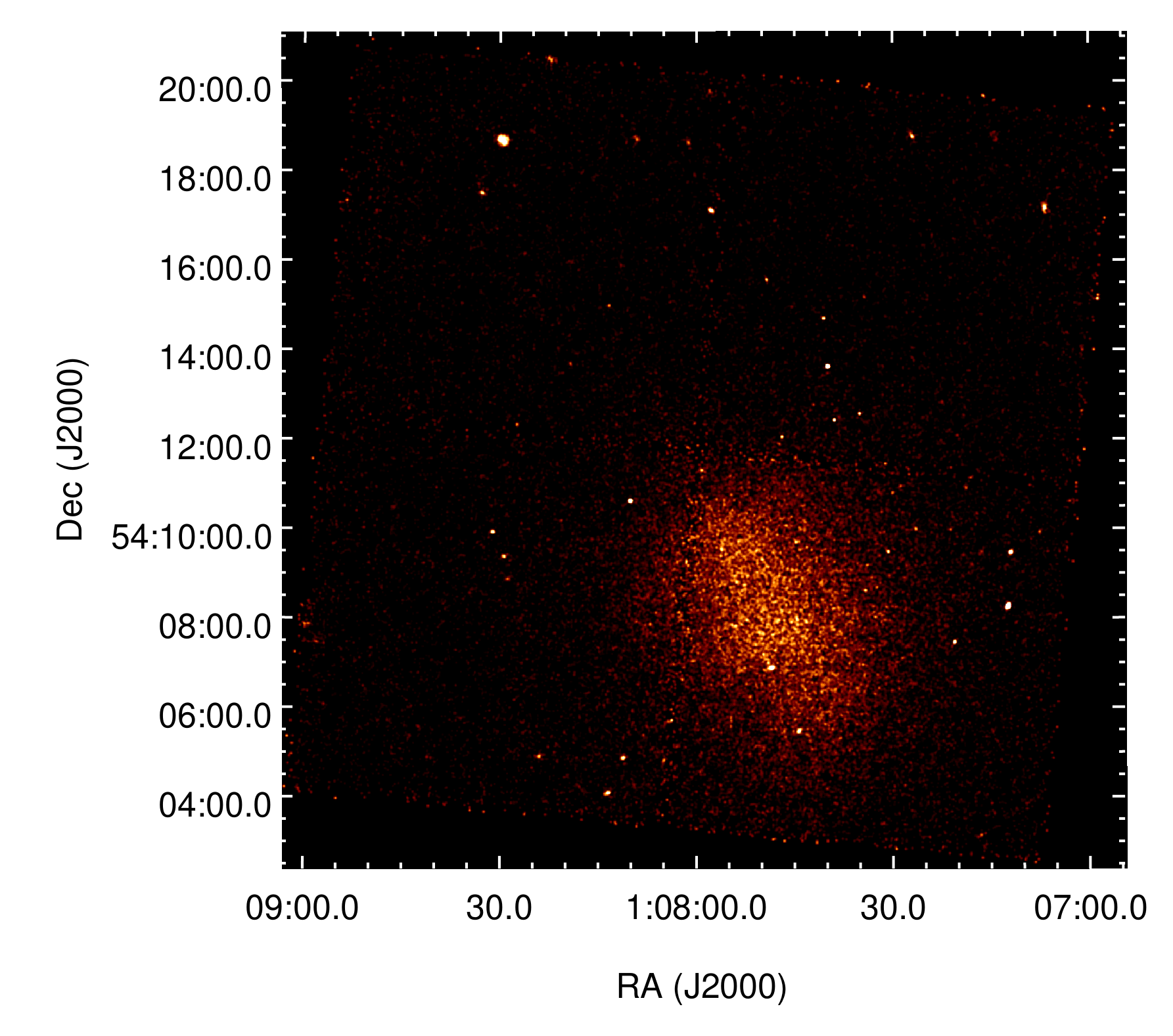}
\vspace{-0.3in}
\caption{
{\it Chandra} 0.3--8.0~keV X-ray image of the full ACIS-I FOV,
background subtracted and 
exposure corrected, and smoothed with a 3\arcsec\ radius Gaussian.
\label{fig:chandra}
}
\vspace{0.03in}
\end{figure}

CIZA~0107 was observed by {\it Chandra} for 23~ks on 2013 June 26
(ObsID~15152).
The aimpoint was on the front-side illuminated ACIS-I CCD.  All data
were reprocessed from the level 1 event files using {\sc CIAO} and
{\sc CALDB~4.6.7}.  CTI and time-dependent gain corrections were
applied. {\sc lc\_clean} was used to check for periods of background
flares\footnote{\url{http://asc.harvard.edu/contrib/maxim/acisbg/}}.
The mean event rate was calculated from a source free region using
time bins within 3$\sigma$ of the overall mean, and bins outside a
factor of 1.2 of this mean were discarded.  There were no periods of
strong background flares, such that the cleaned exposure time was
23~ks.  To model the background we used the {\sc 
  CALDB\footnote{\url{http://cxc.harvard.edu/caldb/}}} 
blank sky background files appropriate for this observation, normalized to match the 10-12~keV count rate in our observations to
account for variations in the particle background. 
Point sources were detected using the {\sc CIAO} tool {\sc wavdetect},
and source regions were checked by eye and subsequently excluded from
the analysis.
The background subtracted, exposure corrected {\it Chandra}
image of the full ACIS-I field of view (FOV) is shown in
Figure~\ref{fig:chandra}.

Unless otherwise specified, all spectra were fitted in the 0.7-8.0~keV
band using {\sc xspec}, with an absorbed {\sc apec} model.  Spectra
were grouped with a minimum of 40 counts per energy bin.
\citet{1998SSRv...85..161G} abundance ratios and {\sc AtomDB~2.0.2} 
\citep{2012ApJ...756..128F} were used throughout.

\subsection{WSRT Radio Observations} \label{sec:wrst}

We use the 1.4 GHz WSRT radio observations presented in
\cite{2011A&A...533A..35V}. The 1.4 GHz image of the cluster has a
noise levels of 29 $\mu$Jy~beam$^{-1}$ and a resolution of
21\arcsec$\times$17\arcsec.  In addition, we make use of a
lower-resolution image with a resolution of 60\arcsec\ and with
emission from compact sources subtracted to better bring out the
diffuse emission. 
For radio imaging, we used a Briggs weighting scheme with a robust
parameter of 0.5 for the full image and natural weighting with a
UV-taper for the source subtracted image.
For more details about these observations and the
data reduction the reader is referred to \cite{2011A&A...533A..35V}.

\subsection{GMRT Radio Observations} \label{sec:gmrt}

On 2009 November 23, CIZA~0107 was observed with the
GMRT \citep{1991ASPC...19..376S} 
simultaneously at 240 and 610~MHz within the scope of a larger project to
study newly discovered radio relics (project code 17\_049; PI van Weeren). The
total time on-source was about 200~minutes. For reducing the data, we used the
SPAM package and a standard data reduction recipe as described in
\citet{2009A&A...501.1185I,2014arXiv1402.4889I}. We used the primary
calibrator 3C48 for deriving 
all instrumental calibrations, adopting the
\citet{2012MNRAS.423L..30S} flux scale. The 
effective bandwidth used for imaging was 6.5 and 32.8~MHz at 240 and 610~MHz,
respectively. The final images were made at resolutions of
13.3\arcsec$\times$10.4\arcsec\ and 5.7\arcsec$\times$4.1\arcsec,
respectively. The achieved 
background RMS noises as measured in the center of the beam were 0.57 and
0.054 mJy~beam$^{-1}$, respectively. We complemented these targeted observations with
a 150~MHz image made from archival
TGSS\footnote{\url{http://tgss.ncra.tifr.res.in/}} survey data (project code
18\_031, pointing R04D65), reprocessed by us in exactly the same way as
described above. Using 16~MHz of bandwidth, and using 3C147 as our primary
calibrator, we obtained an image with resolution of
29.5\arcsec$\times$21.9\arcsec, and a local background RMS noise near the
target of 13~mJy~beam$^{-1}$.
All GMRT images were created using Briggs weighting
  with a robust parameter of -1 to achieve a more stable PSF.

\subsection{VLA Radio Observations}

We use the 73.8 MHz Very Large Array Low-frequency Sky Survey redux
\citep[VLSSr,][]{2014MNRAS.440..327L} to measure the low frequency
emission in CIZA 0107.  
We used the standard VLSSr
  image, which has a spatial resolution of 75\arcsec\ and a measured
noise level of $\sigma_{\rm rms}=78$~mJy~beam$^{-1}$ in the field
around CIZA 0107. We refer the reader to \citet{2014MNRAS.440..327L}
for details of the image calibration and processing.  All flux
measurements include a ``CLEAN bias'' of 0.66$\sigma_{\rm
  rms}$~beam$^{-1}$ for all regions with fluxes above 3$\sigma_{\rm
  rms}$, following the estimates of \citet{2014MNRAS.440..327L}.

\subsection{Optical Observations} \label{sec:optical}

Optical images of CIZA~0107 were taken with the Wide Field Camera
(WFC) on the 2.5~m Isaac Newton Telescope (INT) in the $V$, $R$, and $I$
bands. For more details on the data reduction see
\citet{2011A&A...527A.114V}. We compute galaxy iso-density contours
by counting the number of galaxies per unit area on the sky. For this we
first created a catalog of objects with {\sc sextractor}
\citep{1996A&AS..117..393B}. We excluded all unresolved objects from
the catalog to avoid
the numerous foreground stars in the field. To select cluster members
we only counted objects with  $V-R$ and $R-I$ colors within 0.5
magnitude from the average cD colors ($V-R = 1.0$, $R-I=0.75$). In
addition, we required these objects to be fainter than the cDs but
brighter than magnitude 22 to reduce contamination from more distant
galaxies.  This selection gave 220 cluster member galaxies in the
region spanned by the galaxy density contours shown in
Figure~\ref{fig:optical}.

\section{The Structure of the ICM}

\subsection{Global ICM Properties} \label{sec:global}

We measured the global properties of the ICM by fitting the
X-ray spectrum extracted from within a 645~kpc radius with an absorbed {\sc apec}
model.  
As a CIZA cluster \citep[Clusters in the Zone of
Avoidance,][]{2002ApJ...580..774E,2007ApJ...662..224K}, CIZA~0107 is
at a relatively low Galactic latitude of $b \approx -8.6583$\mydeg.
Due to the relatively high 
absorption close to the Galactic plane, we allowed $N_{\rm H}$ to vary as a
free parameter.  This provided an acceptable fit, with  
$N_{\rm H} = 3.5^{+0.1}_{-0.1}\times 10^{21}$~cm$^{-2}$, $kT =
7.8^{+0.4}_{-0.3}$~keV, and $Z = 0.31^{+0.05}_{-0.05} \, Z_\odot$,
and with a $\chi^2$ per degree of freedom of $\chi^2_\nu = 390/378 =
1.03$.  This fitted value for $N_{\rm H}$ is significantly larger than the
weighted average value from the Leiden/Argentine/Bonn (LAB) survey
\citep{2005A&A...440..775K}
of $N_{\rm H} = 2.3\times 10^{21}$~cm$^{-2}$.  We find that fixing $N_{\rm H}$ at
this value gives a much worse fit, with $kT =
11.9^{+0.6}_{-0.6}$~keV and $\chi^2_\nu = 493/379 = 1.30$.  We note
that there is some systematic uncertainty at low energies associated
with modeling the contamination build-up on the ACIS detector, which
will in principle affect our absorption and temperature measurements,
however this effect is relatively small, $\la10$\%\footnote{http://cxc.harvard.edu/ciao/why/acisqecontam.html}.  We conclude that $N_{\rm H}$ can vary significantly (by at least
50\% or so) on angular scales smaller than what is resolved in the LAB
survey, (which has a resolution of about 0.6\mydeg, larger than the
{\it Chandra} ACIS-I FOV of 0.27\mydeg\ and much larger than {\it
  Chandra's} resolution of $\sim$0.5\arcsec), consistent with results
from some more recent, higher angular resolution \ion{H}{1} survey maps
\citep[e.g.,][]{2011ApJS..194...20P}. 
\begin{figure}
\plotone{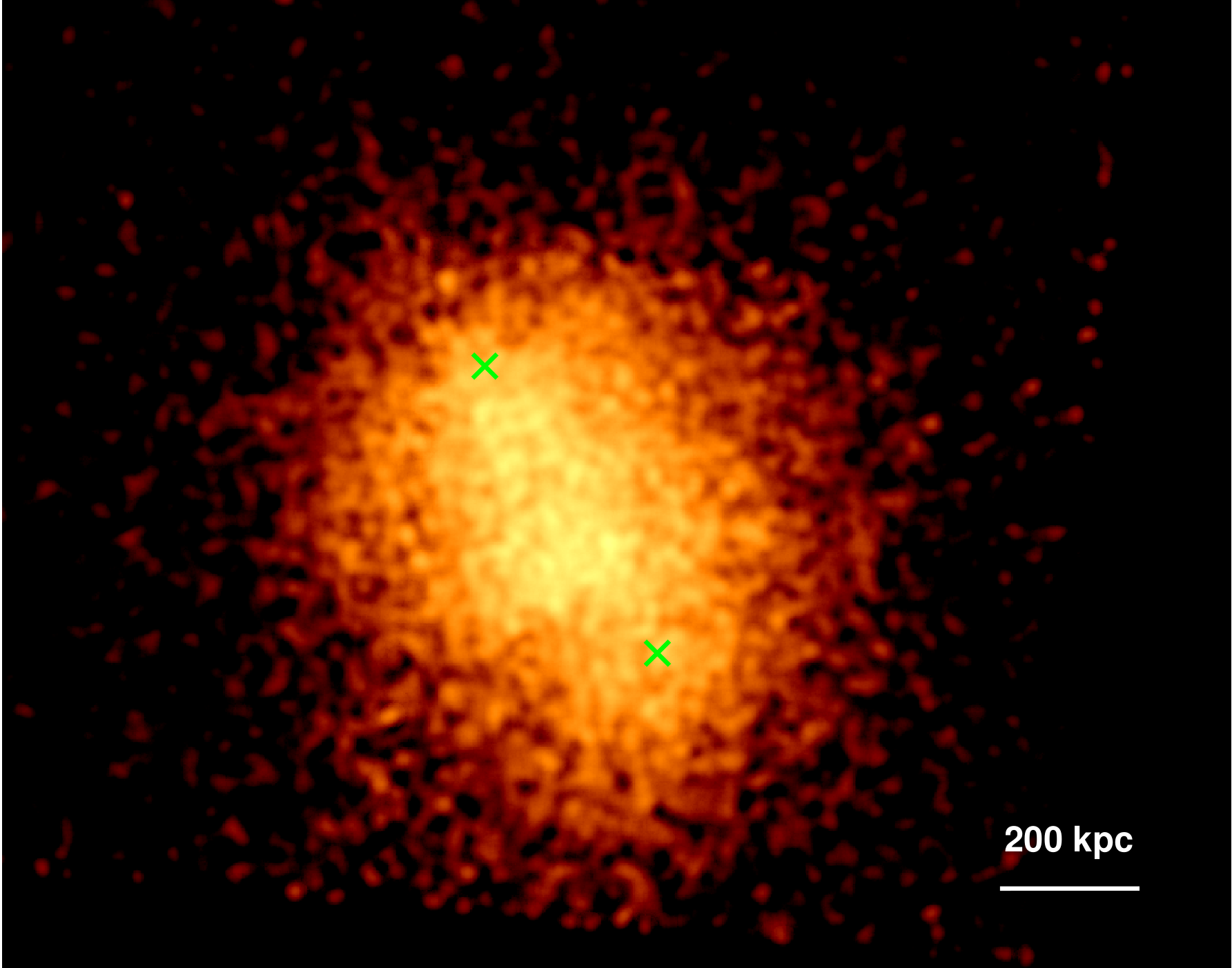}
\caption{
{\it Chandra} 0.3--8.0~keV X-ray image, background subtracted and
exposure corrected, and smoothed with a 10\arcsec\ radius Gaussian.
Point sources have been removed (see text).
The positions of the two BCGs are marked with green crosses.
The ICM is elongated from NE to SW, and the two BCGs are offset from the
X-ray peak along the same axis.
\label{fig:chandra_smo}
}
\end{figure}

We estimated $r_{\rm 500}$, the radius within which the mean density
is equal to 500 times the critical density at the cluster redshift,
and $M_{\rm 500}$, the total mass within $r_{\rm 500}$, using the
$M_{\rm 500}-T_X$ scaling relation of \cite{2009ApJ...692.1033V}.
The radius was iteratively determined by estimating $r_{\rm 500}$,
measuring the temperature by fitting a spectrum extracted with $r <
0.15 \, r_{\rm 500}$ excluded, determining $M_{\rm 500}$ (and $r_{\rm
  500}$) from the $M_{\rm 500}-T_X$ relation, and repeating until the
values converged.  We find $T_{\rm 500} = 7.8^{+0.5}_{-0.5}$~keV,
$M_{\rm 500} = 7.8^{+0.8}_{-0.7} \times 10^{14}$~M$_\odot$, and 
$r_{\rm 500} = 1.35^{+0.05}_{-0.04}$~Mpc.  We note that these
uncertainty ranges
only include statistical uncertainties.  Systematic
uncertainties will likely increase the total uncertainty.
Furthermore, cluster mergers can temporarily boost the global
temperature, leading to biased temperature measurements and inferred masses
\citep{2001ApJ...561..621R,2002ApJ...577..579R}.
Our derived value for $M_{\rm 500}$ is consistent with the
value derived from {\it Planck} Sunyaev–-Zel'dovich observations of
$M_{\rm 500, SZ} = 5.8^{+0.3}_{-0.3} \times 10^{14}$~M$_\odot$ within
3$\sigma$ \citep[{\it Planck} ID PSZ2~G125.37-08.67;][]{2014A&A...571A..29P}.

\subsection{The X-ray Image} \label{sec:image}

\begin{figure}
\plotone{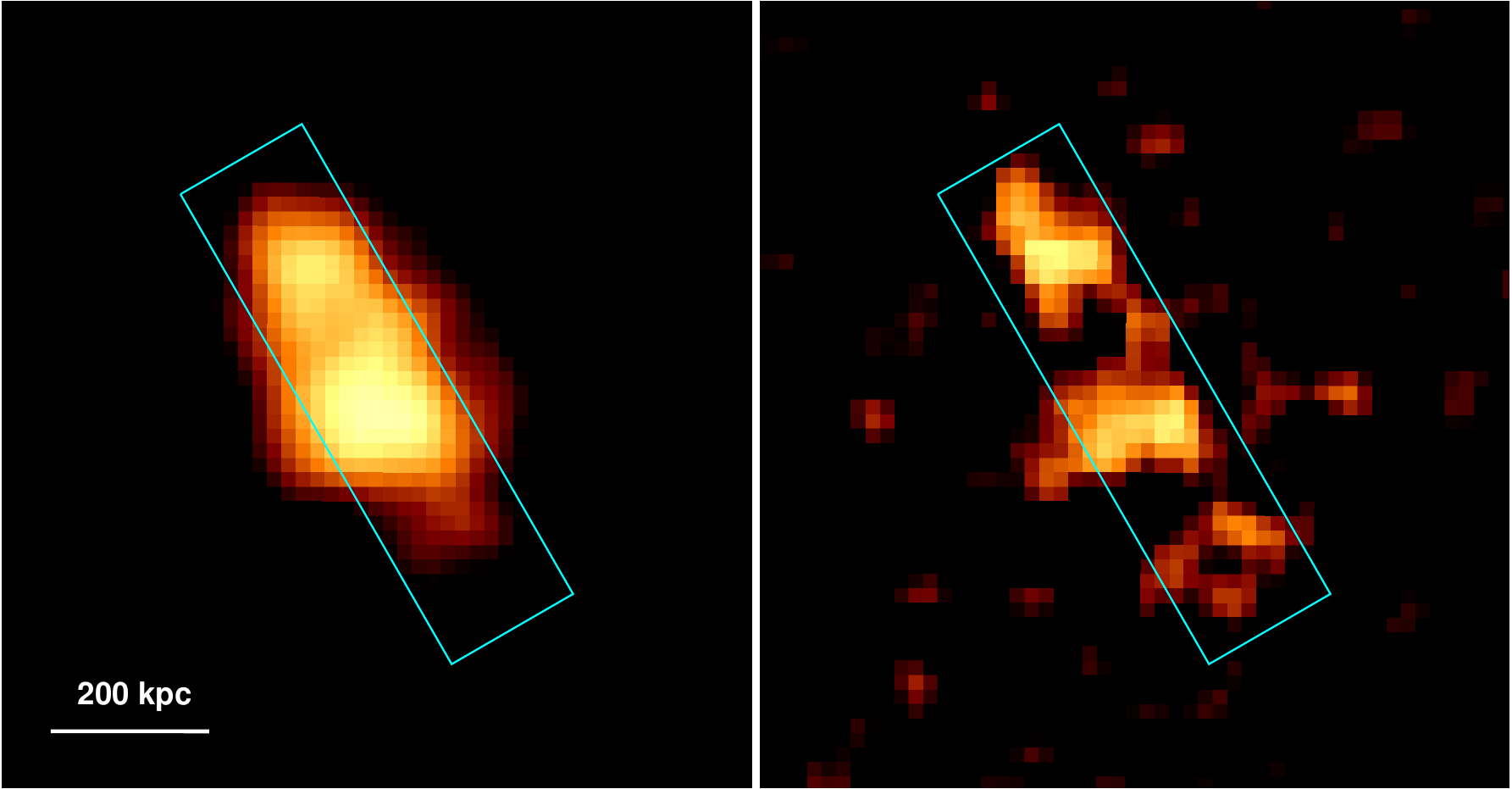}
\caption{
Left: {\it Chandra} image of the core, binned to 10\arcsec pixels and
smoothed with a 30\arcsec\ radius Gaussian, with the color-map and
scale chosen to highlight the bright central bar.  Right:
Unsharp-masked image of the same field.  The light blue rectangular
region shows the
area used to extract the surface brightness profile shown in
Figure~\ref{fig:bar_prof}.
There are two prominent peaks, which we identify as the two merging
subcluster cores, and possibly a fainter peak to the SW, which may be
an enhancement due to a merger bow shock.
\label{fig:core}
}
\end{figure}

The smoothed {\it Chandra} image of CIZA~0107 is shown in
Figure~\ref{fig:chandra_smo}.  Point sources have been excised, and
the resulting gaps filled in by sampling from Poisson distributions
matched to local annular background regions.  The ICM shows a clear
elongation from NE to SW, and no very bright peaks to suggest the presence
of a cool core, indicating a dynamically unrelaxed merging system.
The central bright X-ray emission is non-circular, and also elongated
from NE to SW.  A close up view of this ``central bar'' is shown in
Figure~\ref{fig:core}.  The morphology is suggestive of two distinct
peaks, with a possible faint extension to the SW.

\begin{figure}
\plotone{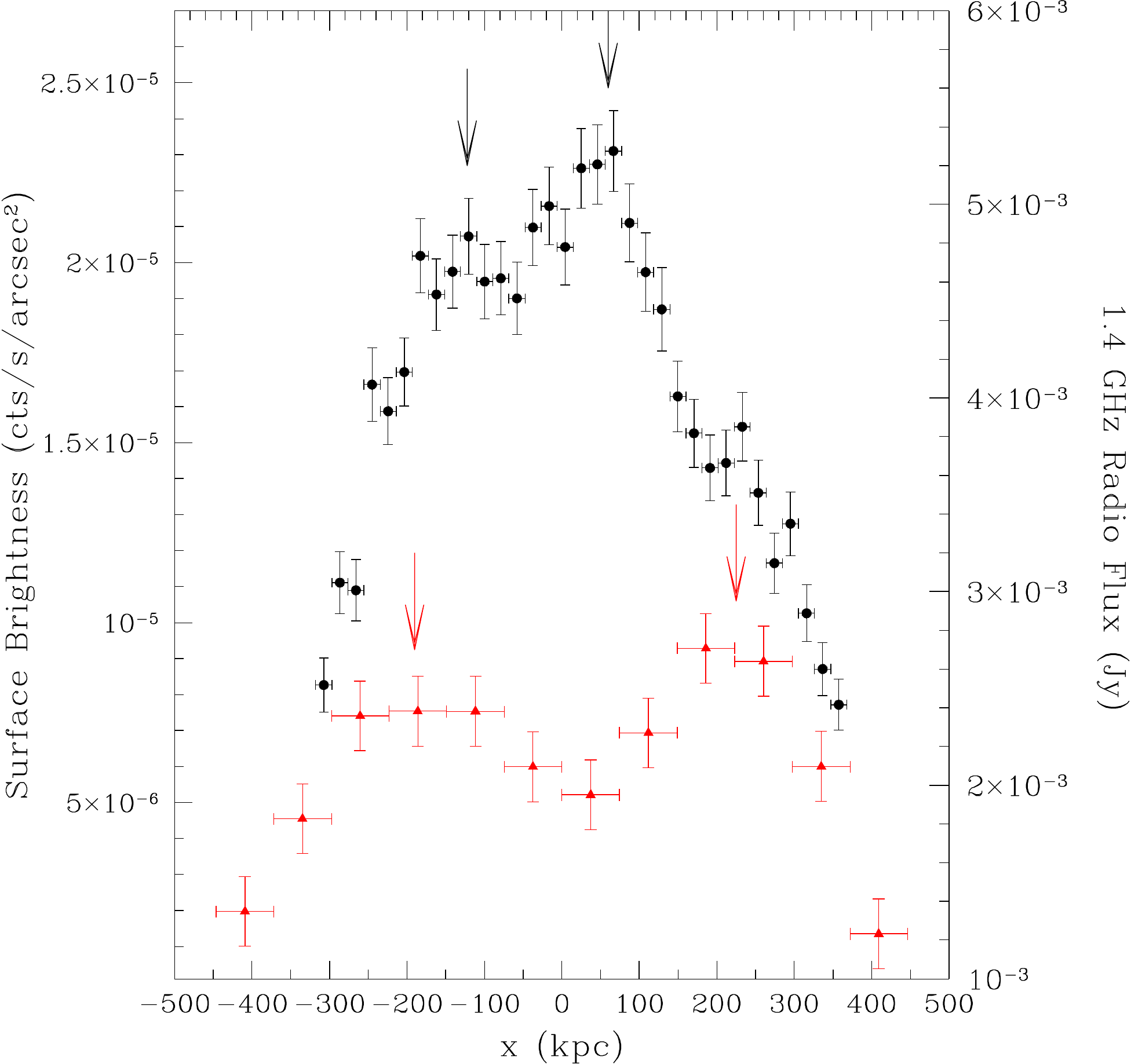}
\caption{
Black circles show the 0.3--8~keV X-ray surface brightness profile
across the box region shown in Figure~\ref{fig:core} (from an
unsmoothed image), while red
triangles show the 1.4~GHz radio flux across the box shown in
Figure~\ref{fig:radio}, from NE to SW.  $x=0$ corresponds to the center
used for the radial profile regions shown in
Figure~\ref{fig:profiles}.  
The instrumental resolution is significantly smaller than the size of
the individual regions in each case.
The black and red arrows indicate the locations of the X-ray and radio
peaks, respectively.
The two radio peaks are farther out than the two X-ray peaks.  There
is some evidence for a weak excess in X-ray surface brightness at the
location of the SW radio peak, possibly due to ICM compression by a
merger bow shock.
\label{fig:bar_prof}
}
\end{figure}

To better show subtle structure in the central bar, we created the
unsharp-masked image shown in Figure~\ref{fig:core}.  This image was
derived by separately smoothing the X-ray image with 10\arcsec and 30\arcsec\
radius Gaussians and taking the difference of the results
\citep[see][]{2006MNRAS.366..417F}.
The two bright peaks are clearly visible, as is a third peak to the
SW.   We
extracted the surface brightness profile across the box region in
Figure~\ref{fig:core}, from NE to SW.  The result is shown in
Figure~\ref{fig:bar_prof}.  The bright central peak is visible at
$\sim 50$~kpc, and the fainter NE peak at $\sim -125$~kpc.  Given the
errors, the NE peak may also be consistent with a flattened tail.
There is a hint of the faint SW peak between 200-300~kpc, although
this feature is not statistically significant.

\subsection{Thermal Structure} \label{sec:tmaps}

To map the thermal structure of the ICM, we generated smoothed
spectral maps using the method described in
\cite{2008ApJ...688..208R}.  
 Spectra were extracted from circular regions, centered on each
 temperature map pixel and
 containing $\sim$2000 source counts in the 0.7-8.0~keV band, and
fitted with an absorbed {\sc apec} model.  Galactic absorption was fixed
at the global value derived in Section~\ref{sec:global}.  Pixel values were
derived from the fitted model parameters.  For comparison, we also
constructed a temperature map using the contour
binning method developed by \cite{2006MNRAS.371..829S}, where
extraction regions are defined based on surface brightness contours,
using a signal-to-noise (S/N) ratio of 44 per region.

\begin{figure*}
\plotfour{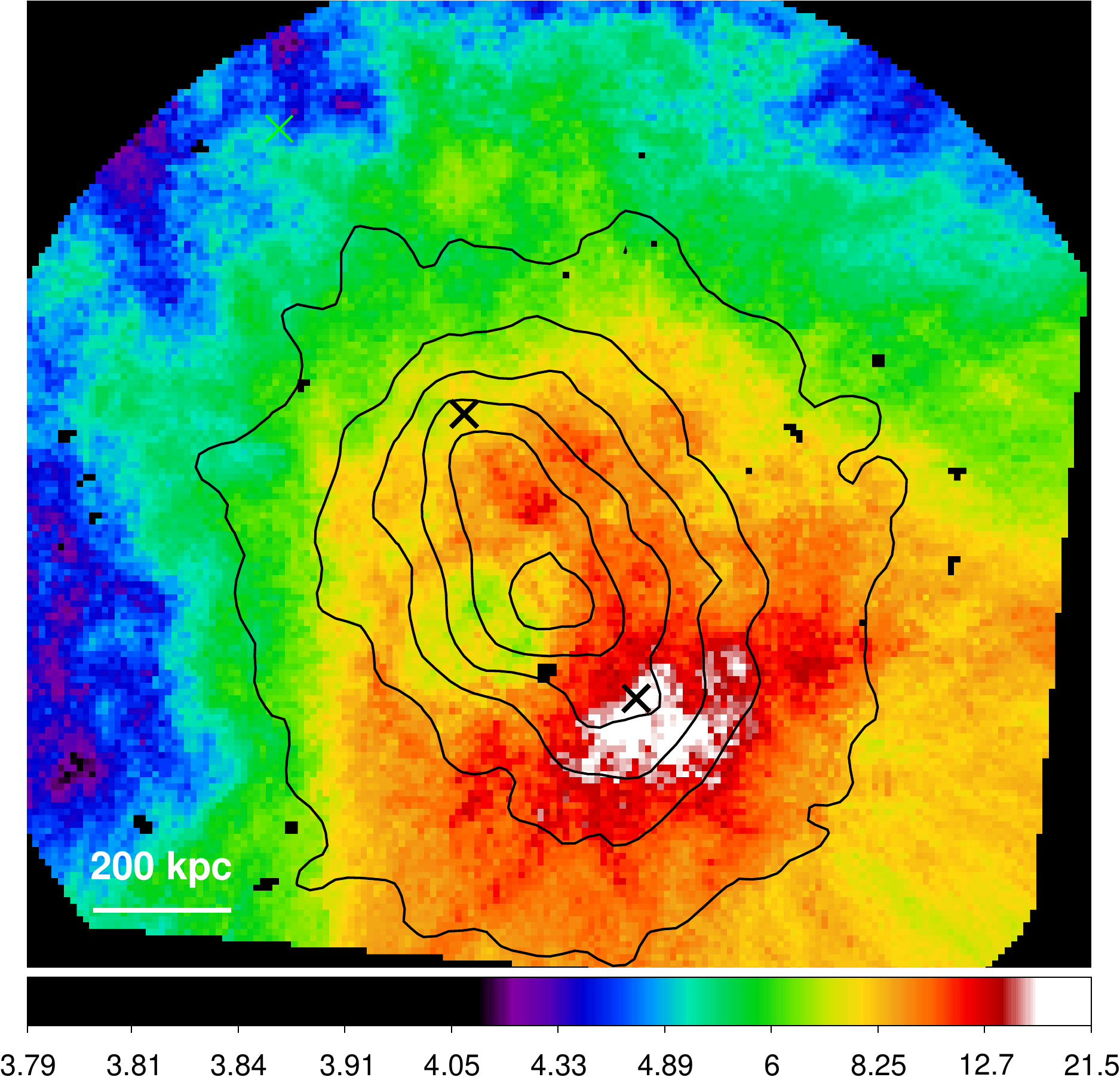}{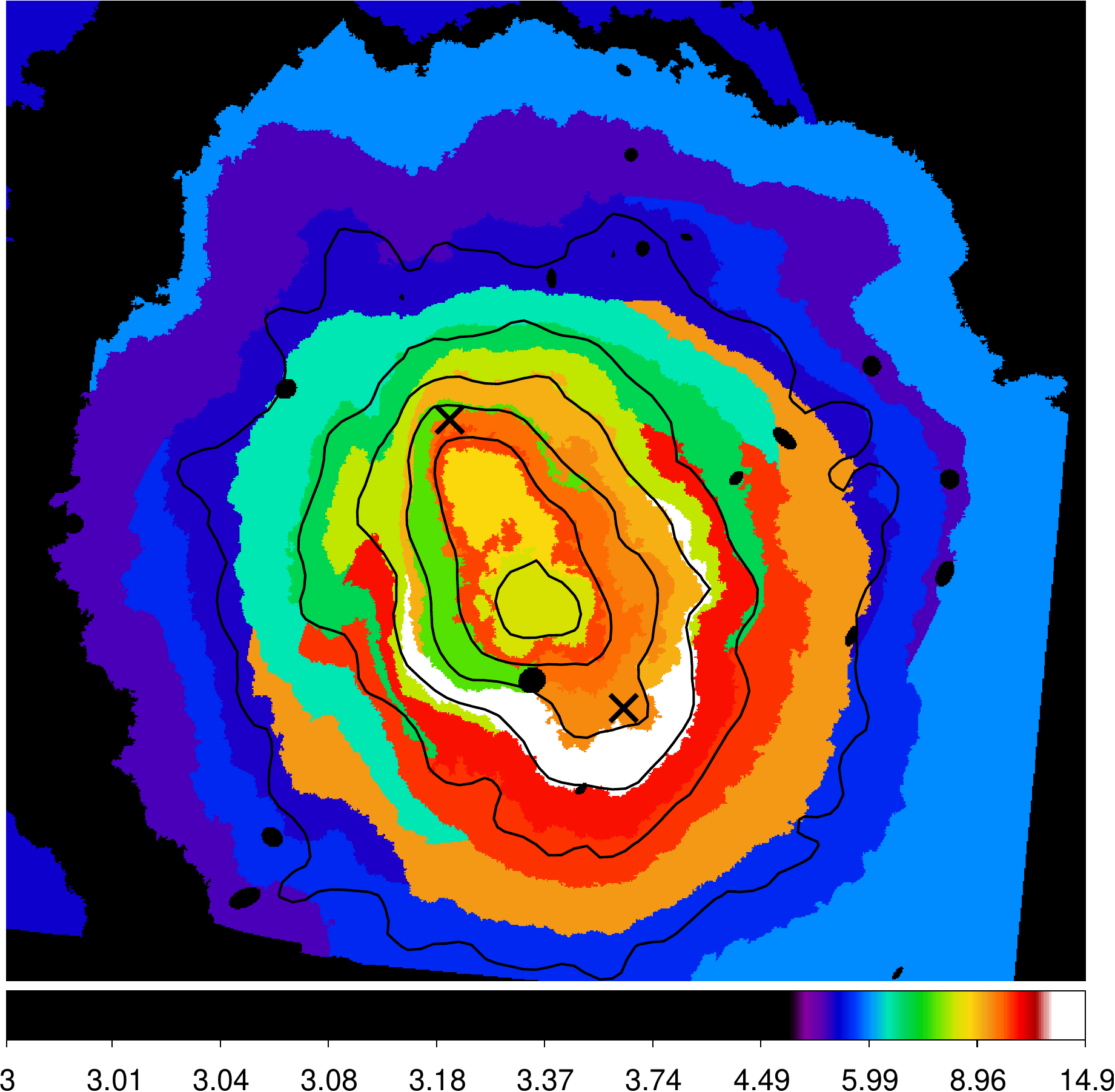}{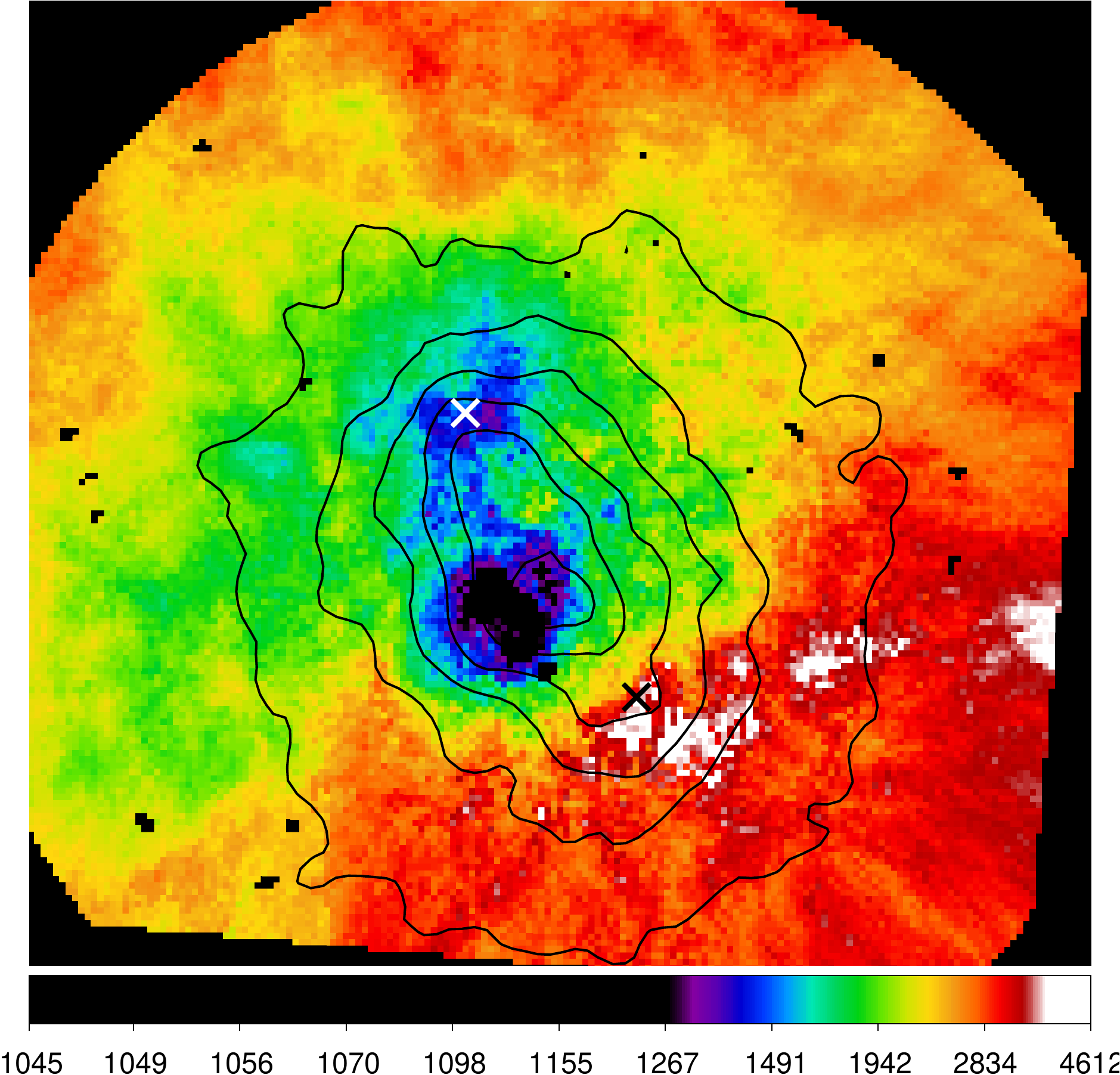}{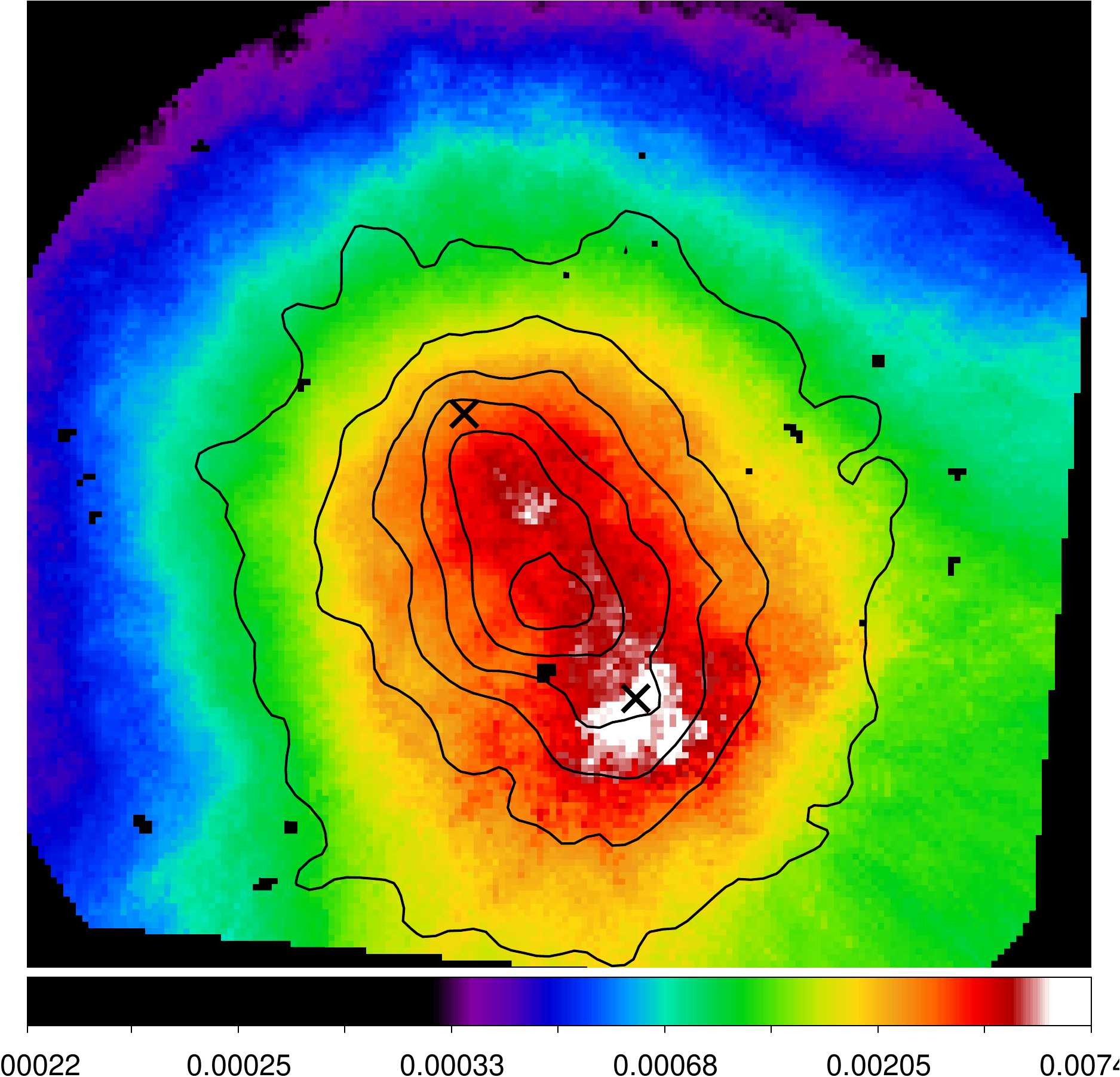}
\caption{Left: Smoothed temperature map, in units of keV, with the
  {\it Chandra} surface brightness contours overlaid in black.
  Left-center: Contour binned temperature map of the same region.
  Right-center: Smoothed pseudo-entropy map, in arbitrary units,
  calculated as $kT A^{-1/3}$, where $A$ is the {\sc apec} normalization scaled
  by the area of the extraction region (blue is low pseudo-entropy).
  The lowest entropy gas is extended from NE to SW, consistent with
  two partially disrupted, merging cluster cores.
  Right: Smoothed pseudo-pressure map, in arbitrary units, calculated as $kT
  A^{1/2}$ (white is high pseudo-pressure).  There is a pressure peak
  to the SW, coincident with the high temperature peak, consistent
  with the presence of a merger shock.  The crosses mark the locations
  of the two BCGs in each panel.
\label{fig:tmaps}
}
\end{figure*}

The {\it Chandra} temperature, pseudo-entropy, and pseudo-pressure
maps are shown in Figure~\ref{fig:tmaps}.  Both the smoothed
and the contour binned temperature maps show a high temperature region
to the SW, beyond the X-ray surface brightness peaks, and in the region of
the SW optical galaxy density peak and the bright SW radio relic
(Section~\ref{sec:radio}).
This region appears as a pressure peak in the pseudo-pressure map.
The pseudo-entropy map shows a central low entropy region that is
extended from NE to SW, consistent with the presence of two partially
disrupted, merging cluster cores. 
Thus, the spectral maps support a roughly head-on merger scenario with
a NE-SW merger axis, with two subcluster cores and the presence of
high-temperature, shock-heated gas to the SW.

\begin{figure*}
\plotone{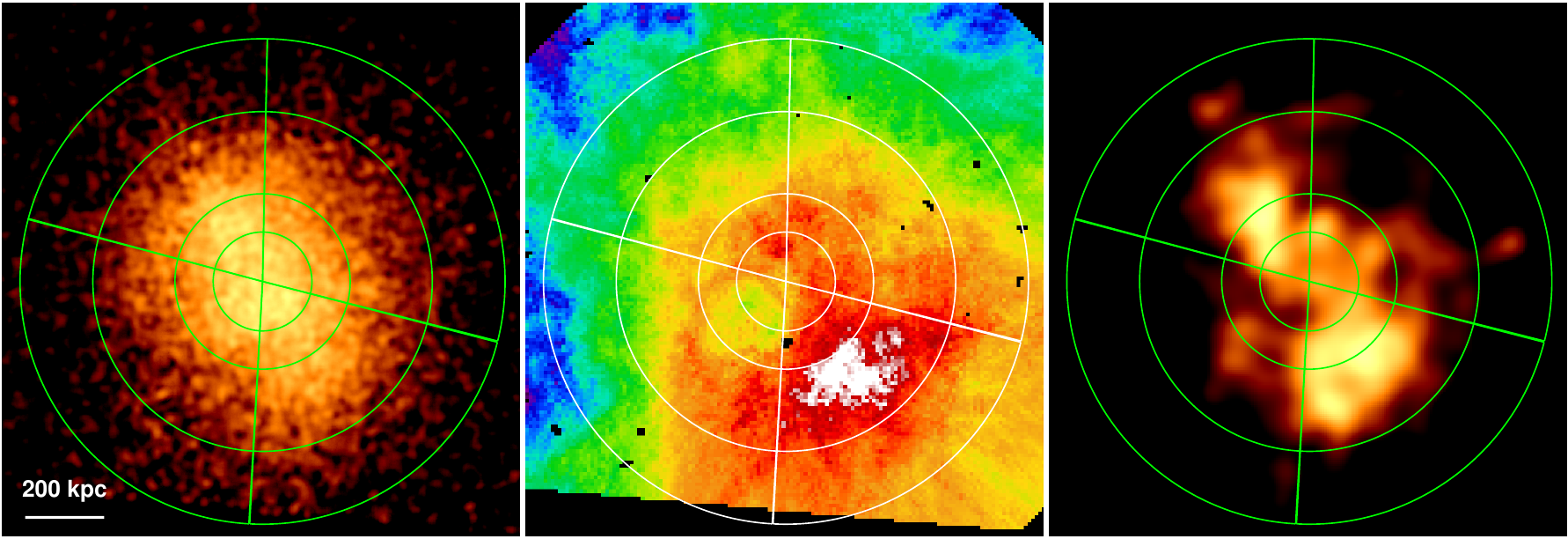}
\caption{X-ray image (left), smoothed temperature map (center), and
  1.4~GHz radio image (right) with the extraction regions used
  to make the profiles shown in Figure~\ref{fig:ktprof} overlaid.
\label{fig:profiles}
}
\end{figure*}

\begin{figure}
\plotone{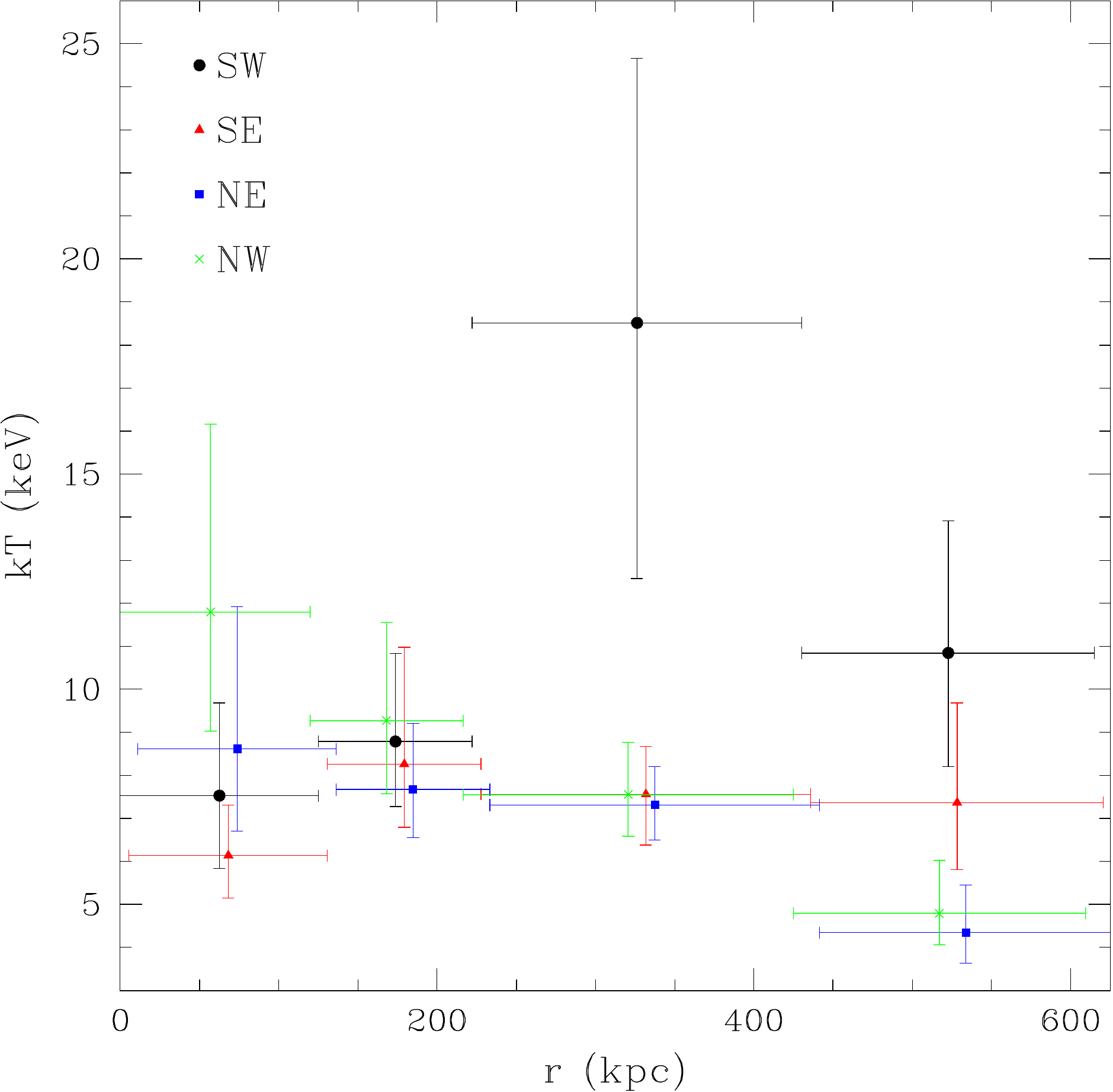}
\caption{
Temperature profile in the four sectors shown in Figure~\ref{fig:profiles}.
Individual profiles have been shifted slightly in the radial direction
for clarity.
\label{fig:ktprof}
}
\end{figure}

To determine the significance of the SW temperature peak, we measured
the projected temperature profile in four directions (see the angular sectors
shown in Figure~\ref{fig:profiles}).  Due to the limited number of
counts, we chose only four bins per sector.  
We note that the amount of absorption can affect temperature
measurements, with underestimated $N_{\rm H}$ biasing the temperature high.
Since $N_{\rm H}$ is relatively
large in this region of the sky, and may change across the field, the
absorption was
allowed to vary in each fit.  The abundance was also allowed to vary.
The resulting temperature profiles are shown in
Figure~\ref{fig:ktprof}.  
The highest temperature by far is to the SW,
in the region of the temperature peak seen in the temperature map,
with  $kT = 18.5 \pm 6.0$~keV.  
Compared with other sectors in the same radial range, this feature is
significant at about 1.8$\sigma$ (see Section~\ref{sec:dynamical}).
Since $N_{\rm H}$ is allowed to vary, its uncertainty is folded into the
temperature uncertainty.  If we fix $N_{\rm H}$ at our best fitting global
ICM value of $3.5\times 10^{21}$~cm$^{-2}$, we find a statistically
equivalent fit with $kT = 14.6^{+2.4}_{-3.1}$~keV for the SW
temperature peak.  
In this case, the temperature increase is significant at 2.3$\sigma$
as compared with other sectors in the same radial range also with $N_{\rm
  H}$ fixed at the global value.
If we instead fix $N_{\rm H}$ at the LAB survey value of
$2.3\times 10^{21}$~cm$^{-2}$ we find a somewhat worse statistical fit
(with a null hypothesis probability of 2.8\% versus 5.1\% for the free
$N_{\rm H}$ case) with $kT = 30^{+9}_{-6}$~keV.  For every annular bin, the
best-fitting value of $N_{\rm H}$ is within 2$\sigma$ of the best-fitting
global ICM value.
Although the SW temperature peak is only marginally statistically
significant, based on its location on the merger axis and correlation
with the SW radio relic, as expected for a merger bow shock (see
Section~\ref{sec:dynamical}), we conclude that it is likely a real feature.

We note
that, given the high temperature, lack of strong emission lines, and
limited number of counts, we
cannot confirm the thermal nature of this emission (as compared with,
e.g., inverse Compton emission from the interaction of the CMB with
the radio emitting particles in this region, or emission from
unresolved point sources associated with the nearby BCG).  
This issue is discussed further in Section~\ref{sec:ic}.

\section{Diffuse Radio Emission} \label{sec:radio}

\begin{figure*}
\plotone{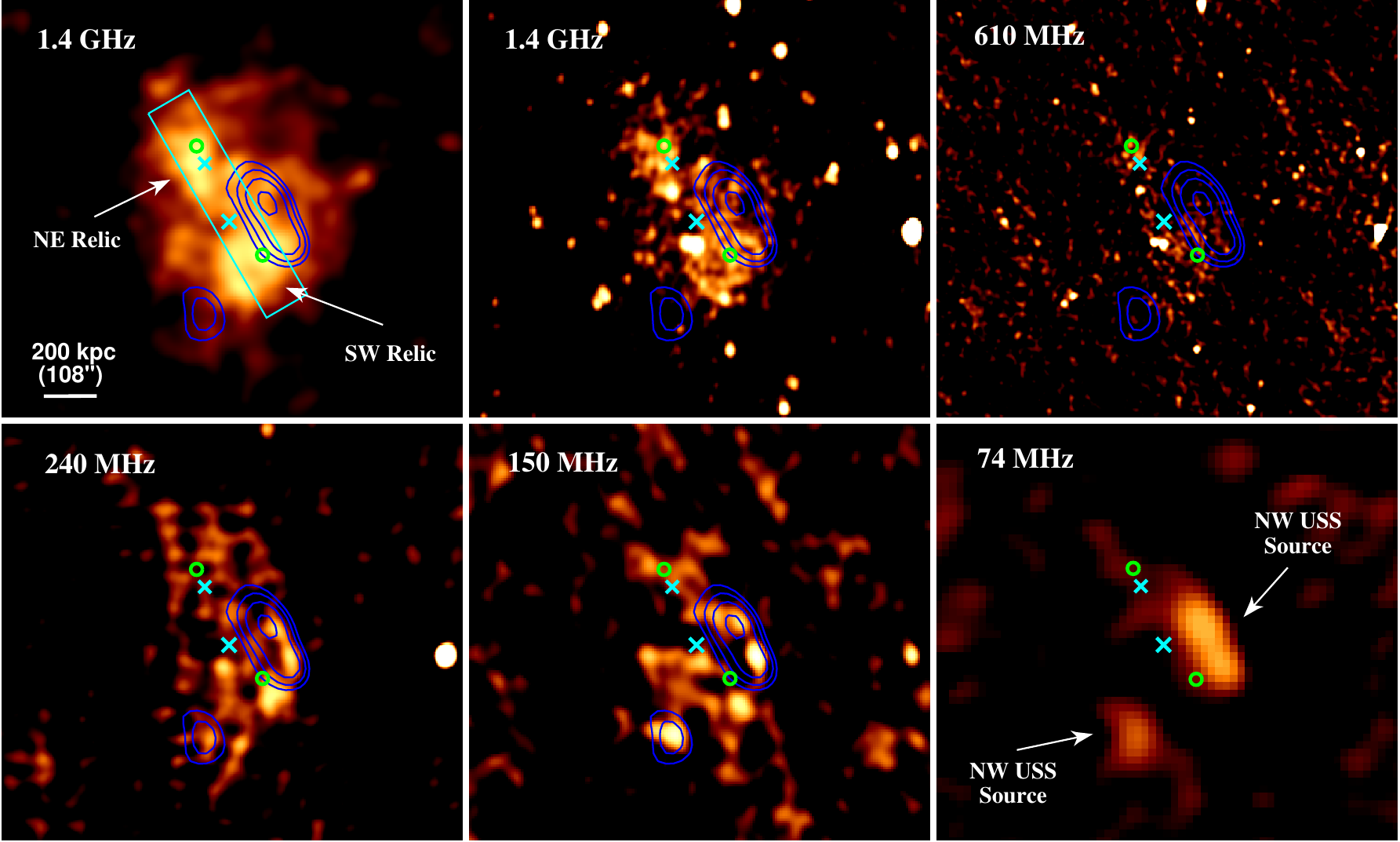}
\caption{Top left: 1.4~GHz WSRT image (with a resolution of
  21\arcsec$\times$17\arcsec\ and a noise level of
  29~$\mu$Jy~beam$^{-1}$), with 
  point sources removed, and 
  smoothed with a 40\arcsec\ radius Gaussian to highlight faint structure.
  The cyan box indicates the region used to extract the radio and
  X-ray profiles
  shown in Figure~\ref{fig:bar_prof}. Top-middle: 1.4~GHz WSRT
  image with point source included.  Top right: 610~MHz GMRT image
  (with a resolution of 5.7\arcsec$\times$4.1\arcsec\ and a noise
  level of 54~$\mu$Jy~beam$^{-1}$),
  smoothed with a 9\arcsec\ radius Gaussian.  Bottom left: 240~MHz GMRT
  image
  (with a resolution of 13.3\arcsec$\times$10.4\arcsec and a noise
  level of 0.57~mJy~beam$^{-1}$), smoothed with a 24\arcsec\ radius
  Gaussian.  Bottom center:
  150~MHz GMRT image
  (with a resolution of 29.5\arcsec$\times$21.9\arcsec\ and a noise
  level of 13~mJy~beam$^{-1}$), smoothed with a 25\arcsec\ radius Gaussian.
  Bottom right: 73.8~MHz NRAO VLA VLSSr image (with a resolution of
  75\arcsec\ and a noise level of 78~mJy~beam$^{-1}$).  
  The 73.8~MHz VLSS contours are shown in blue, they start at 3$\sigma$
  (0.653~Jy~beam$^{-1}$) and increase by factors of $\sqrt{2}$.
  The locations of the BCGs are indicated by green circles, and the
  locations of the two X-ray peaks shown in Figure~\ref{fig:bar_prof} are
  indicated by cyan crosses.
  The double radio relic and USS
  sources have apparently uncorrelated morphologies.  The relic is
  most clearly visible at 1.4~GHz, and is no longer detected at
  74~MHz, while the USS sources are bright at 74~MHz, are somewhat
  visible up to 240~MHz, and are no longer detected at 1.4~GHz.
\label{fig:radio}
}
\end{figure*}

The radio emission is shown in Figure~\ref{fig:radio}, ranging
from 74~MHz to 1.4~GHz.  In \cite{2011A&A...533A..35V}, the diffuse
1.4~GHz radio emission was classified as a radio halo with a
north-south extension, based on the continuous structure of the
source and the rough correspondence with the low resolution {\it
  ROSAT} X-ray image.  The new {\it Chandra} images reveal a NE-SW
extension of the ICM, with a doubly peaked core (Section~\ref{sec:image}).
In light of these new data, we suggest that the diffuse radio emission
represents a two-sided radio relic, possibly with radio halo emission
between the relics, as has been observed in other
systems
\citep[e.g.,][]{2006Sci...314..791B,2011A&A...528A..38V,2012MNRAS.425L..36V,2014MNRAS.444.3130D},
although other interpretations are possible.  For a detailed
discussion, see Section~\ref{sec:relic}.
For the purposes of this paper, we refer to the diffuse 1.4~GHz
emission as a radio relic throughout.

The 1.4~GHz radio flux profile extracted from
the box region shown in Figure~\ref{fig:radio} is compared with the
X-ray surface brightness profile in the same region in
Figure~\ref{fig:bar_prof}.  
The radio profile shows two clear peaks (in contrast with the single
central peak expected for radio halos), which are at larger
radii than the two bright X-ray peaks, in the regions of the BCGs and
optical galaxy density peaks.  The SW radio peak is roughly
coincident with the SW temperature peak identified in Section~\ref{sec:tmaps}.
The observations are therefore consistent with what is expected for radio relics
in a late stage merging system, where the relics trace ICM merger shocks,
which lead the X-ray cores.  The dynamical state of CIZA~0107 is
discussed further in Section~\ref{sec:dynamical}.

\begin{figure}
\plotone{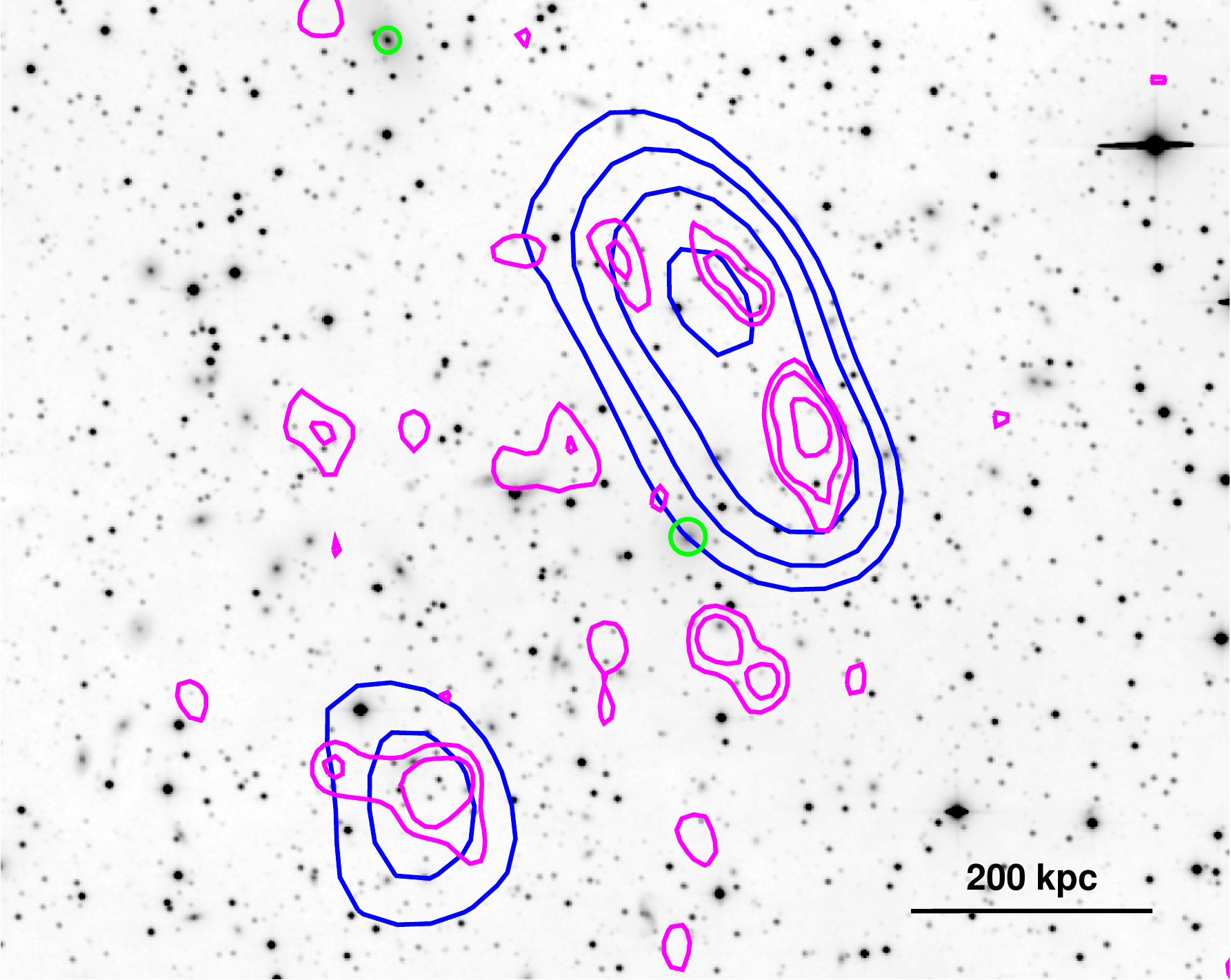}
\caption{Optical INT $I$~band image in the region for the USS radio sources.
  74~MHz VLSSr contours are overlaid in blue, and 150~MHz GMRT
  contours are overlaid in magenta.  Contours start at 3$\sigma$ at
  74~MHz and 1$\sigma$ at 150~MHz
  and increase by factors of $\sqrt{2}$.  The two BCGs are marked with
  green circles.  There are no obvious optical point sources clearly
  associated with the USS diffuse radio sources.
\label{fig:optical_radio}
}
\end{figure}

The 74~MHz image reveals an elongated
structure, NW of the southern subcluster center, and a fainter secondary peak to
the SE (Figure~\ref{fig:radio}, lower right panel and blue contours).
The NW feature is seen as an
extended, possibly double peaked source at 150~MHz and 240~MHz, but is
not visible at 610~MHz or 1.4~GHz (see the contours in
Figure~\ref{fig:radio}).  It is not coincident with the brighter
regions of the relic at 1.4~GHz, suggesting that it may be an
unrelated feature.  
Similarly, the SE 74~MHz peak is seen at 150~MHz, and possibly
240~MHz, but not at higher frequencies. Neither feature is clearly
associated with a bright X-ray or optical source (see
Figure~\ref{fig:optical_radio}).

In Table~\ref{tab:radio}, we give radio fluxes at 1.4~GHz, 240~MHz,
150~MHz, and 74~MHz, along with some spectral indices, in three
different regions: a larger region containing all of the diffuse
emission detected at all frequencies, a smaller region corresponding
to the NW USS source at 74~MHz, and a slightly smaller region
corresponding to the SW USS source at 74~MHz.  
Fluxes were not calculated at 610~MHz since we expect some flux loss
on the scale of the diffuse emission due to the lack of short
baselines.
To account for calibration uncertainties, we include a systematic
error of 
5\% at 1.4~GHz, 10\% at 240~MHz and 150~MHz, and 15\% at 74~MHz.
We follow the calibration uncertainties adopted by
\citet{2014MNRAS.440..327L} for the VLSSr data.
For the WSRT and the GMRT, our adopted calibration uncertainties are
based on our experience with
  working with data from these observatories, as we have done
  elsewhere \citep[e.g.,][]{2014ApJ...781L..32V}.
For the total emission in the region of the relic, we find
spectral indices from 
74~MHz to 150~MHz and 74~MHz to 240~MHz of
$\alpha^{150}_{74} = -1.2$ and $\alpha^{240}_{74} = -1.4$.  For the NW, elongated
structure detected at 74~MHz (Figure~\ref{fig:radio}), we find
significantly steeper emission, with $\alpha^{150}_{74} = -2.3$ and
$\alpha^{240}_{74} = -2.1$.  Similarly, for the SE peak at 74~MHz we
find  $\alpha^{150}_{74} = -2.2$ and $\alpha^{240}_{74} = -2.1$ (see
Figure~\ref{fig:radio_spec}). These 
features are discussed further in Section~\ref{sec:uss}.

\begin{deluxetable*}{lcccccc}
\tablewidth{0pt}
\tablecaption{Radio Source Properties \label{tab:radio}}
\tablehead{
\colhead{Source}&
\colhead{$F_{1400}$}&
\colhead{$F_{240}$}&
\colhead{$F_{150}$}&
\colhead{$F_{74}$}&
\colhead{$\alpha^{240}_{74}$}&
\colhead{$\alpha^{150}_{74}$}\\
\colhead{}&
\colhead{(mJy)}&
\colhead{(mJy)}&
\colhead{(mJy)}&
\colhead{(mJy)}&
\colhead{}&
\colhead{}
}
\startdata
Total&$72\pm4.5$&$329\pm41$&$759\pm163$&$1779\tablenotemark{a}\pm637$&$-1.4\pm0.3$&$-1.2\pm0.6$\\
NW USS Source&$29\pm1.9$&$174\pm21$&$420\pm80$&$2126\pm412$&$-2.1\pm0.2$&$-2.3\pm0.4$\\
SE USS Source&$6.8\pm0.9$&$66\pm10$&$159\pm50$&$778\pm223$&$-2.1\pm0.3$&$-2.2\pm0.6$
\enddata
\tablenotetext{a}{We note that the 74~MHz flux in the total region is
  less than the sum of the fluxes in the NW and SE USS regions, which
are contained within the total region.  This is because the local
noise level in the region of the cluster is biased negative due to
imperfect calibration.  The total and summed NW plus SE 74~MHz fluxes
are consistent within 1.5$\sigma$.}
\end{deluxetable*}

\section{Discussion} \label{sec:discuss}

\subsection{Cluster Dynamical State} \label{sec:dynamical}

X-ray and radio observations of CIZA~0107 show that it is a
non-dynamically relaxed merger system, with the merger axis likely
along the NE-SW direction.  The overall X-ray emission is extended
along this axis, with two  X-ray surface brightness peaks
along the same line, consistent with two merging subclusters.
There is a high temperature region to the SW, along the
merger axis, consistent with the presence of a merger bow shock
leading the SW core.  
There is a hint of an excess in the X-ray surface brightness at this
location (Figures~\ref{fig:core}~\&~\ref{fig:bar_prof}),
consistent with the presence of a shock, although this feature is
only marginally significant.
The radio emission is also extended from NE to
SW, at multiple frequencies, with two bright peaks at 1.4~GHz roughly
200~kpc from the overall cluster center (the center of the annular bin
regions shown in Figure~\ref{fig:profiles}). The SW radio peak is
coincident with the SW shock region.  This is
consistent with a double radio relic, as seen in other merging
systems, where particles in the ICM are (re)accelerated at the merger
shock associated with each subcluster and emit synchrotron radiation.
\begin{figure}
\plotone{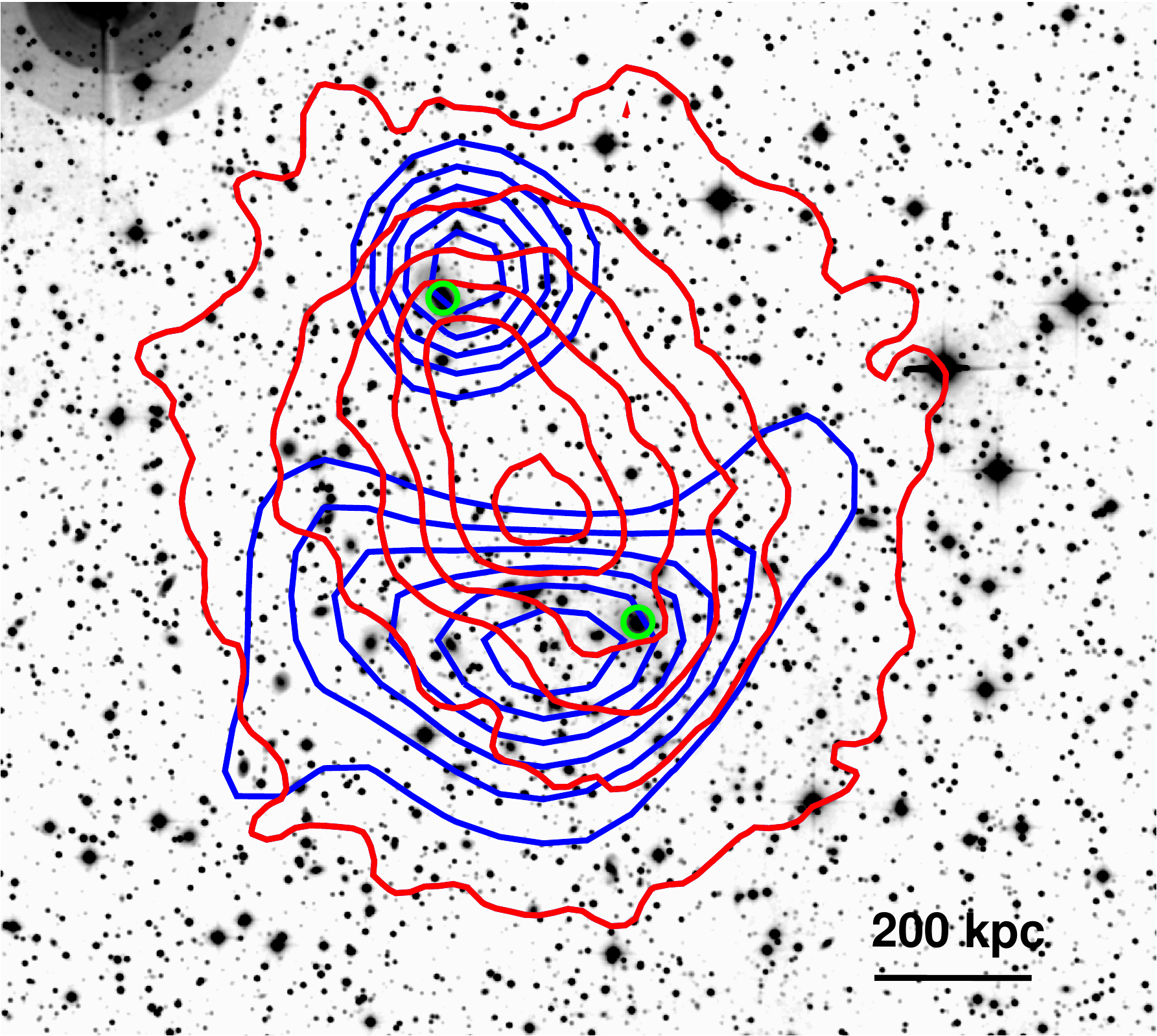}
\caption{
Optical INT $I$~band image with {\it Chandra} X-ray contours
overlaid in red; smoothed, source flux weighted galaxy number density
contours overlaid in blue; and the locations of the BCGs indicated with green
circles.  The two optical galaxy density peaks are clearly offset from
the peak of the X-ray emission along the merger axis, from NE to SW.
\label{fig:optical}
}
\end{figure}

The optical image is shown in Figure~\ref{fig:optical}, with the {\it
  Chandra} X-ray surface brightness contours overlaid in red, and the
optical flux-weighted cluster galaxy density contours overlaid in blue.
This image shows two BCGs, indicated with green circles, consistent
with two merging subclusters.  The SW BCG
was identified by \cite{1995MNRAS.274...75C}, with a reported redshift
of $z=0.109$ (although this redshift is described as ``provisional''
due to poor S/N). 
Both of the optical galaxy density peaks and both
BCGs lie on the merger axis, at larger radii than the X-ray peaks,
and separated from one another by roughly 500~kpc in projection.  We
conclude that CIZA~0107 is likely a dissociative post-merger system, similar
to the famous Bullet cluster \citep{2002ApJ...567L..27M,
  2007PhR...443....1M}.

In this scenario, the diffuse ICM in the subcluster cores experiences
ram pressure drag forces during the cluster merger, while the
effectively collisionless galaxies and dark matter (DM) halos do not,
leading to a separation
between the ICM and galaxy/DM peaks.  Therefore, the ICM is expected
to trail the galaxies roughly along the merger axis \citep[although the gas
may lead the galaxies during the later stages of a merger due to the
ram pressure slingshot effect, see][]{2007PhR...443....1M}.
The ICM
and galaxy density peak offsets indicate that this is a post core
passage merging system.
Thus we expect the NE X-ray and galaxy peaks to be moving NE, and
their SW counterparts to the SW (in projection).
As in the Bullet cluster, the SW galaxy density peak is at roughly the
location of the SW leading bow shock.
The lack of complicated structure in the ICM along with the subcluster
cores, the merger shock, the radio relics, the optical galaxy density
peaks, and the BCGs all lying roughly along the same line (i.e., the
merger axis) suggest
that this is a low impact parameter merger (at least in projection),
also similar to the Bullet cluster.

Figure~\ref{fig:ktprof} shows a temperature increase
between 200-400~kpc to the SW as compared with other sectors, which
all have $kT \approx 7.5$~keV.  Averaging the NW, NE, and SE sectors
together in this radial range, we find a temperature of $kT = 7.5 \pm
0.6$~keV, as compared with $kT = 18.5 \pm 6.0$~keV in the SW.
Applying the standard Rankine-Hugoniot shock jump conditions for an
ideal gas with a constant ratio of specific heats of $\gamma = 5/3$,
this corresponds to a Mach number of $M = 2.3 \pm 0.4$.  In principle,
this is a lower limit on the true Mach number, since projection
effects will tend to bias the measured temperature increase low.
For a 7.5~keV gas, this Mach number corresponds to a relative ICM velocity of
3250~\kms.  As shown by \cite{2007MNRAS.380..911S}, the Mach number of
an ICM merger shock can overestimate the relative velocity of the
subcluster DM halos, as is likely the case for the Bullet cluster.
\begin{figure*}
\plottwo{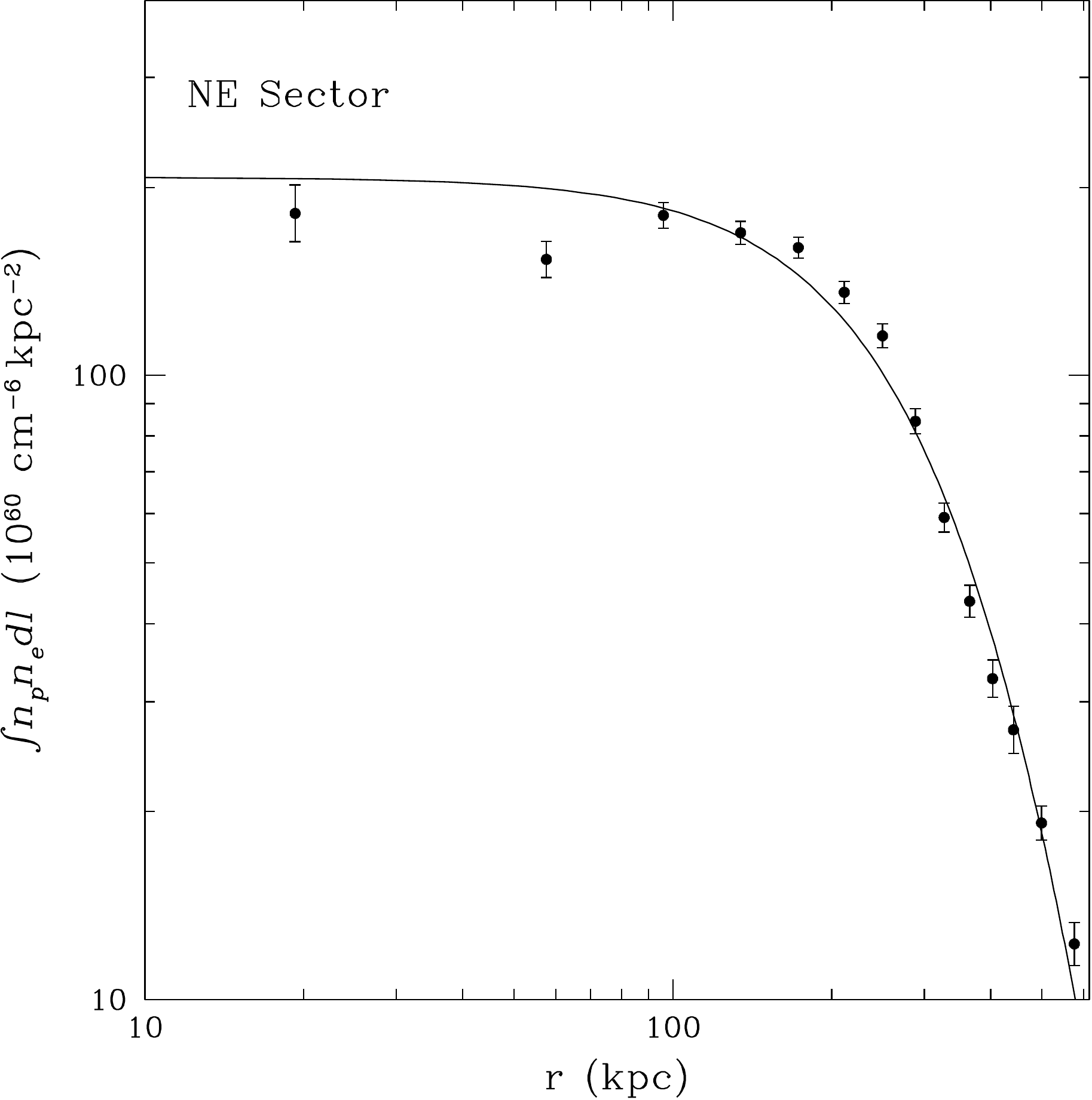}{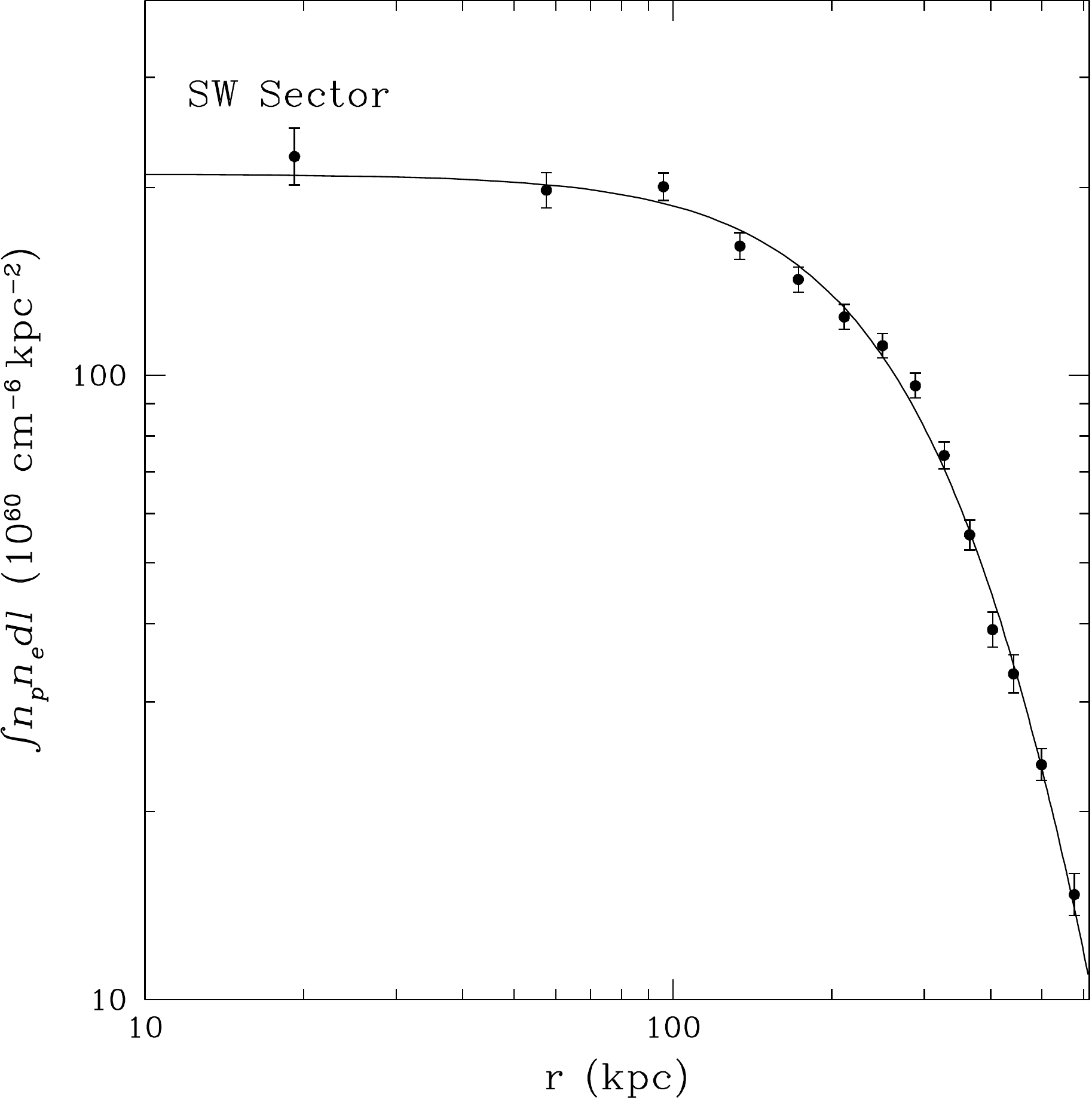}
\caption{Integrated emissivity profiles in the SW and NE sectors.  The
  solid lines show the best-fitting $\beta$-models.  $n_p$ and $n_e$
  are the proton and electron number densities, respectively, and the
  integral is taken along the line of sight $l$.
\label{fig:emprofs}
}
\end{figure*}

When viewed along a tangent line to the Mach cone, ICM shocks are
expected to appear as sharp surface brightness edges in high angular
resolution X-ray observations.  However, no such edge is clearly
visible in the region of the SW shock, neither in the X-ray image nor
the unsharp-masked image.  To further test for an edge in this region,
we extracted the integrated emissivity profiles in the SW and NE
sectors.  These profiles were generated assuming the radial
temperature profile in each sector followed the projected temperature
profile (Figure~\ref{fig:ktprof}) with the abundance fixed at $Z = 0.3
\, Z_\odot$.  The results are shown fitted with a 3D $\beta$-model
density profile in Figure~\ref{fig:emprofs}.  Both sectors are
reasonably well-described by the model, with no clear evidence for an
edge feature, although the SW profile shows weak evidence for a dip in
the profile at $\sim370$~kpc, at the outer edge of the SW radio relic
and high temperature region.

\begin{figure}
\plotone{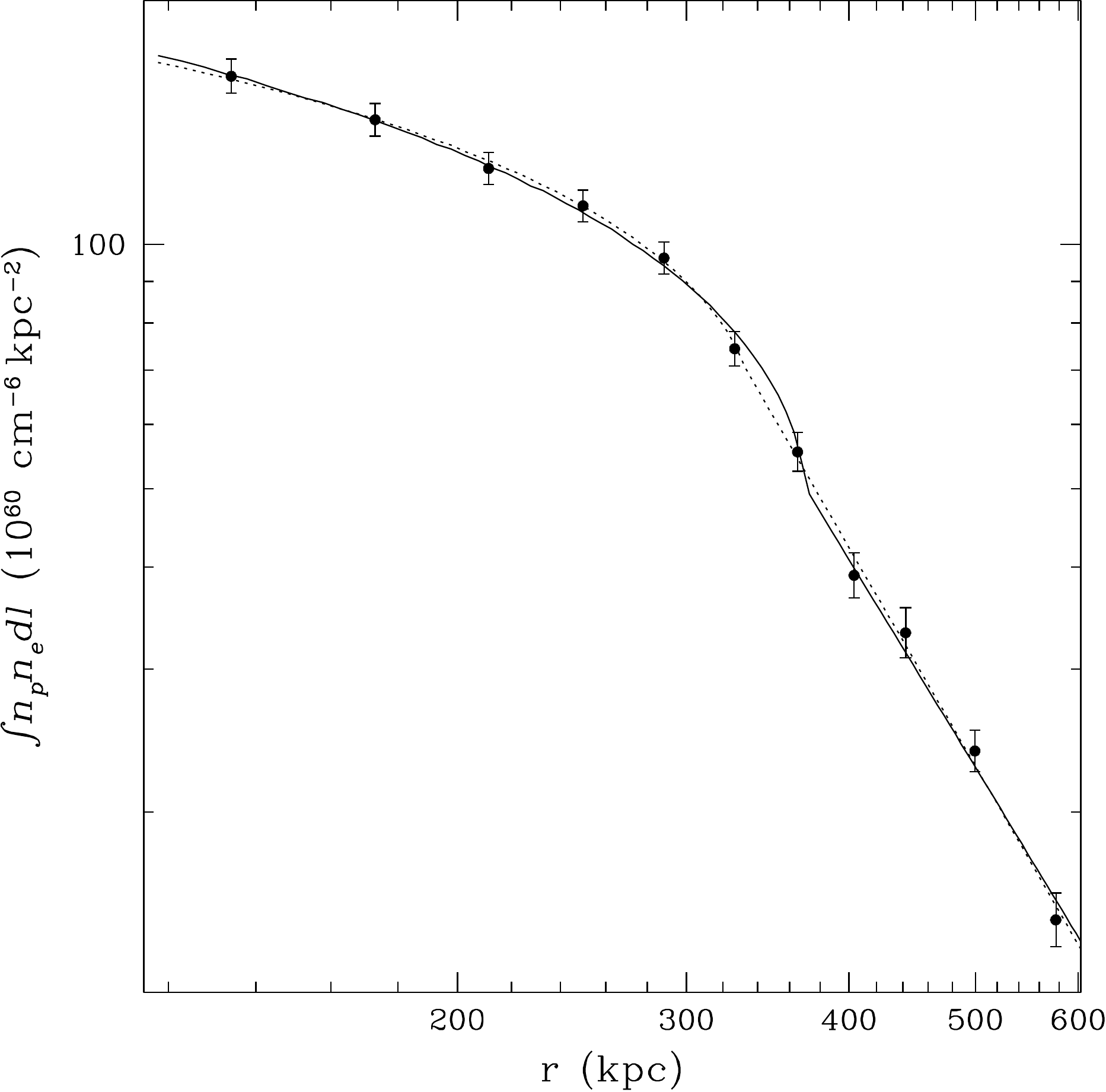}
\caption{
SW IEM profile fit with a discontinuous double power-law density profile with
$r_{\rm break} \equiv 370$~kpc (solid line) and with a continuous
double power-law model (dotted line).
\label{fig:sw_edge}
}
\end{figure}

We modeled this dip using a discontinuous double power-law density
profile of the form
\begin{equation} \label{eq:nr}
n_e(r) = \left\{
\begin{array}{ll}
n_0 ( \frac{r}{r_{\rm br}} )^{-k_1} & (r \le r_{\rm  br})\\
n_1 ( \frac{r}{r_{\rm br}} )^{-k_2} & (r > r_{\rm  br})  \\
\end{array}
\right. ,
\end{equation}
where $r_{\rm br}$ is the break radius.  This model is shown fitted to
the data in Figure~\ref{fig:sw_edge}.  If we fix the break
radius at $r_{\rm br}=370$~kpc, we find evidence for a weak edge, corresponding to
a density jump factor of $1.2^{+0.1}_{-0.1}$.  However, if the
location of the break radius is allowed to vary the profile is also
well-fit by a model with no discontinuous jump and a change in
slope at a smaller radius.  The lack of an apparent sharp
surface brightness edge may indicate that there is a line
of sight component to this merger, such that the edge is
diminished by projection effects, while the hot, shock-heated gas
is still visible in projection, as has been reported for other systems,
(e.g., \citealp[A2443,][]{2013ApJ...772...84C};
\citealp[A2744,][]{2011ApJ...728...27O}).
The angle between the merger axis and the plane of the sky needn't be
very large to obfuscate shock edges \citep[see Figure~18 in][]{2011ApJ...728...27O}.
Additionally, it may be difficult to
pick out edges by eye in this relatively shallow (23~ks) observation,
especially in fainter regions where the image must be smoothed to show
ICM structure.
The radius of curvature of merger shock fronts is
generally not expected to be centered on the cluster itself.  Since we
are unable to identify the location of the edge, we cannot match the
radius of curvature, center, and bin boundary location of our profile
extraction regions to the edge.  This will effectively blur the edge
in the extracted profile, making it more difficult to detect.
We conclude that deeper observations are required to confirm or rule
out the presence of a surface brightness edge in this region.

The BCG of each subcluster was observed with the Wide Field Grism Spectrograph 2
on the University of Hawaii 2.2-meter Mauna Kea telescope as part of
the CIZA survey \citep{2002ApJ...580..774E, 2007ApJ...662..224K}. They
find redshifts of 0.103 for both of the BCGs (D. Kocevski private
communication). While the exact uncertainty of these redshifts are not reported,
they add to the supporting X-ray, radio, and optical imaging evidence
that CIZA~0107 is composed of two nearby clusters that are undergoing
a major merger.

\subsection{Dark Matter Self-Interaction Cross Section} \label{sec:sidm}

\cite{2004ApJ...606..819M} outline four methods for placing
constraints on the self-interaction cross-section of DM ($\sigma_{\rm
  DM}$) using
dissociative mergers.
Placing such constraints on $\sigma_{\rm DM}$ requires
lensing observations to map the total mass distribution, as well
as optical spectroscopy to identify subcluster member galaxies, optical
centroids, and subcluster line-of-sight velocities, neither of which
are currently available for CIZA~0107.
Nevertheless, we can use simple estimates to show that this system
can, in principle, provide competitive constraints on $\sigma_{\rm
  DM}$.  To this end, let us assume that the dark matter halos are
centered on the optical galaxy density peaks shown in
Figure~\ref{fig:optical}.  The offsets between the DM peaks and gas
peaks indicates that the scattering depth of the DM particles cannot
be much larger than 1, otherwise the DM subhaloes would experience drag forces
similar to the gas and there would be no offset.  Following
\cite{2004ApJ...606..819M}, we can write the subcluster DM scattering
depth as
\begin{equation} \label{eq:tau}
  \tau_s = \frac{\sigma_{\rm DM}}{m_{\rm DM}} \Sigma_s,
\end{equation}
where $\Sigma_s$ is the DM surface mass density.  To estimate
$\Sigma_s$, we assume that the gas is isothermal 
and follows a single $\beta-$model density profile (which is roughly
consistent with observations, see
Figures~\ref{fig:ktprof}~\&~\ref{fig:emprofs}), and that it is in
hydrostatic equilibrium.  The latter assumption is not strictly true for this
merging system, but suffices for the simple estimate we make here.  In
this case, the total density profile can be written as
\begin{equation}
  \rho(r) = \frac{3 \beta kT}{4 \pi G \mu m_p} \left[ \frac{x^2(3+x^2)}{(1+x^2)^2}  \right],
\end{equation}
where $x \equiv r/r_c$, $r_c$ and $\beta$ are the core radius and
index that describe the gas density profile, $kT$ is the gas
temperature, and $\mu m_p$ is the average mass per gas particle.
Integrating over a circular aperture with projected radius $x_0$ and along the
line of sight, and dividing by the area of the aperture, we find that
the  projected surface mass density is
\begin{equation}
\begin{aligned}
\Sigma_s & = \frac{4}{x_0^2} \int_{0}^{x_0}x \int_{x}^{\infty}\frac{r
  \rho(r)}{\sqrt{r^2 - x^2}} dr \, dx \\ & = \frac{3 \beta kT}{2 G \mu m_p
  \sqrt{x_0^2 + r_c^2}}.
\end{aligned}
\end{equation}
From the IEM profile fits in Section~\ref{sec:dynamical}, we find $\beta
\approx 1.8$ and $r_c \approx 630$~kpc.  For $kT = 8$~keV, this gives
$\Sigma_s \approx 0.26$~g~cm$^{-2}$ within a fiducial projected radius
of $x_0 = 100$~kpc.  From Equation~\ref{eq:tau} we
then find that $\sigma_{\rm DM}/m_{\rm DM} \la 4$~cm$^2$~g$^{-1}$. 
This can be compared with a similar estimate based on observations of
the Bullet cluster, but with $\Sigma_s$ measured from lensing
observations rather than estimated based on scaling relations, which
gives $\sigma_{\rm DM}/m_{\rm DM} \la 5$~cm$^2$~g$^{-1}$
\citep{2004ApJ...606..819M}.

We stress that, given the lack of gravitational lensing observations
to locate the DM peaks, we do not consider our limit on $\sigma_{\rm
  DM}/m_{\rm DM}$ to be meaningful (here, we have assumed that the DM
peaks are coincident with the optical centroids, which would imply $\sigma_{\rm
  DM}=0$).  Rather, the intent is to demonstrate
that this system can potentially provide constraints on $\sigma_{\rm
  DM}/m_{\rm DM}$ that are comparable to those derived for the Bullet
cluster using similar methods, which currently provides the tightest
constraints of any individual system known.  Optical lensing and galaxy
spectroscopy observations of CIZA~0107 will allow more accurate
constraints to be placed on the DM self-interaction cross-section
using a variety of methods, as has been done previously for the Bullet
cluster \citep{2004ApJ...606..819M,2008ApJ...679.1173R}.

\subsection{The Nature of the Diffuse Radio Emission} \label{sec:diffuse}

\subsubsection{High Frequency Diffuse Radio Emission} \label{sec:relic}

The radio image at 1.4~GHz shows two distinct peaks in the diffuse
emission along the merger axis (Section~\ref{sec:radio},
Figure~\ref{fig:radio}).  In principle, each peak may represent a
radio relic associated with a merger shock, or a radio halo associated
with a subcluster core.  Double (as opposed to single) radio relics
are not uncommon \citep[e.g.,][]{2011A&A...528A..38V,
  2014MNRAS.444.3130D}, while double radio halos are extremely rare
but not undocumented \citep[e.g.,][]{2010A&A...509A..86M}.  There
exists an empirical correlation between the radio halo power at
1.4~GHz, $P_{\rm 1.4 GHz}$, and the 0.1--2.4~keV X-ray luminosity
within $r_{\rm 500}$, $L_{X,500}$.  For the entire system, including
both radio peaks, we find $P_{\rm 1.4 GHz} = 1.9 \times
10^{24}$~W~Hz$^{-1}$ and $L_{X,500} = 3.9 \times
10^{44}$~erg~s$^{-1}$.  This places it somewhat above the $P_{\rm 1.4
  GHz}$~--~$L_{X,500}$ relation of \cite{2013ApJ...777..141C},
although it is within the range of the scatter about this relation.
Thus, we can not rule out a double radio halo interpretation of this
based on the radio and X-ray powers.  
We note that we
  can likely exclude an AGN origin for this diffuse emission on Mpc
  scales since the synchrotron loss time is too short for electrons to
have time to diffuse over these large regions (hence, in-situ particle
acceleration is needed).

The SW radio peak is coincident with the high temperature,
presumably shock-heated gas to the SW
(Figures~\ref{fig:tmaps}~\&~\ref{fig:ktprof}).
It is offset from the SW X-ray peak by
$\sim200$~kpc (Figure~\ref{fig:bar_prof}), roughly coincident with the
SW BCG (Figure~\ref{fig:radio}), inconsistent with what is expected
for a radio halo, which should remain centered on the ICM of its host
cluster. 
There is a hint of a small increase in X-ray surface brightness at the
location of the SW radio peak (also seen as the faint peak to the SW
in Figure~\ref{fig:core}), but this feature is too faint to be one of
the main subcluster cores, and is instead likely due to ICM
compression at the shock front.
Given the coincidence with the high temperature
  region, a possible local increase in the X-ray surface brightness,
  its placement along the merger axis, and its separation from the
  X-ray surface brightness peak, we suggest that the SW radio peak may
  be a radio relic that has been energized by a local merger shock.

For the NE radio peak, the interpretation is less clear.  The offset
from the NE X-ray peak is smaller and less significant
(Figure~\ref{fig:bar_prof}), and there is no indication of an X-ray
shock in this region.  Thus, we conclude that this system is most
likely either a
double radio relic, possibly with a faint radio halo between the
relics, or a SW radio relic and NE radio halo.  Due to the relative
rarity of these configurations, the double radio relic interpretation
is probably more likely.  
The somewhat flocculent appearance of the relic, in contrast with the
sharp, linear morphology of some other relics
\citep[e.g.,][]{2012A&A...546A.124V}, could in principle be due to a
small inclination angle between the merger axis and the plane of the
sky, as suggested by the lack of a sharp shock front edge in the X-ray
(Section~\ref{sec:dynamical}).
For simplicity, we refer to the high
frequency radio structure as a double radio relic, or simply ``the
relic'', throughout.

\subsubsection{Ultra-Steep Spectrum Radio Sources} \label{sec:uss}

The 74~MHz radio image
(Figure~\ref{fig:radio}) shows two distinct peaks near the SW
subcluster (an elongated
structure to the NW, and a separate peak to the SE) that are not aligned
with the merger axis, nor do they correlate with the brighter 1.4~GHz
emission.  The peaks become less prominent with increasing frequency, and are no
longer visible at $\nu \ga 610$~MHz.  This is a reflection of the very
steep spectral indices of these sources, with $\alpha_{\rm NE}
\approx \alpha_{\rm SW} \approx -2$ (see Section~\ref{sec:radio}).
Neither source is clearly associated with a distinct radio, optical,
or X-ray source (Figure~\ref{fig:optical_radio}).

\begin{figure}
\plotone{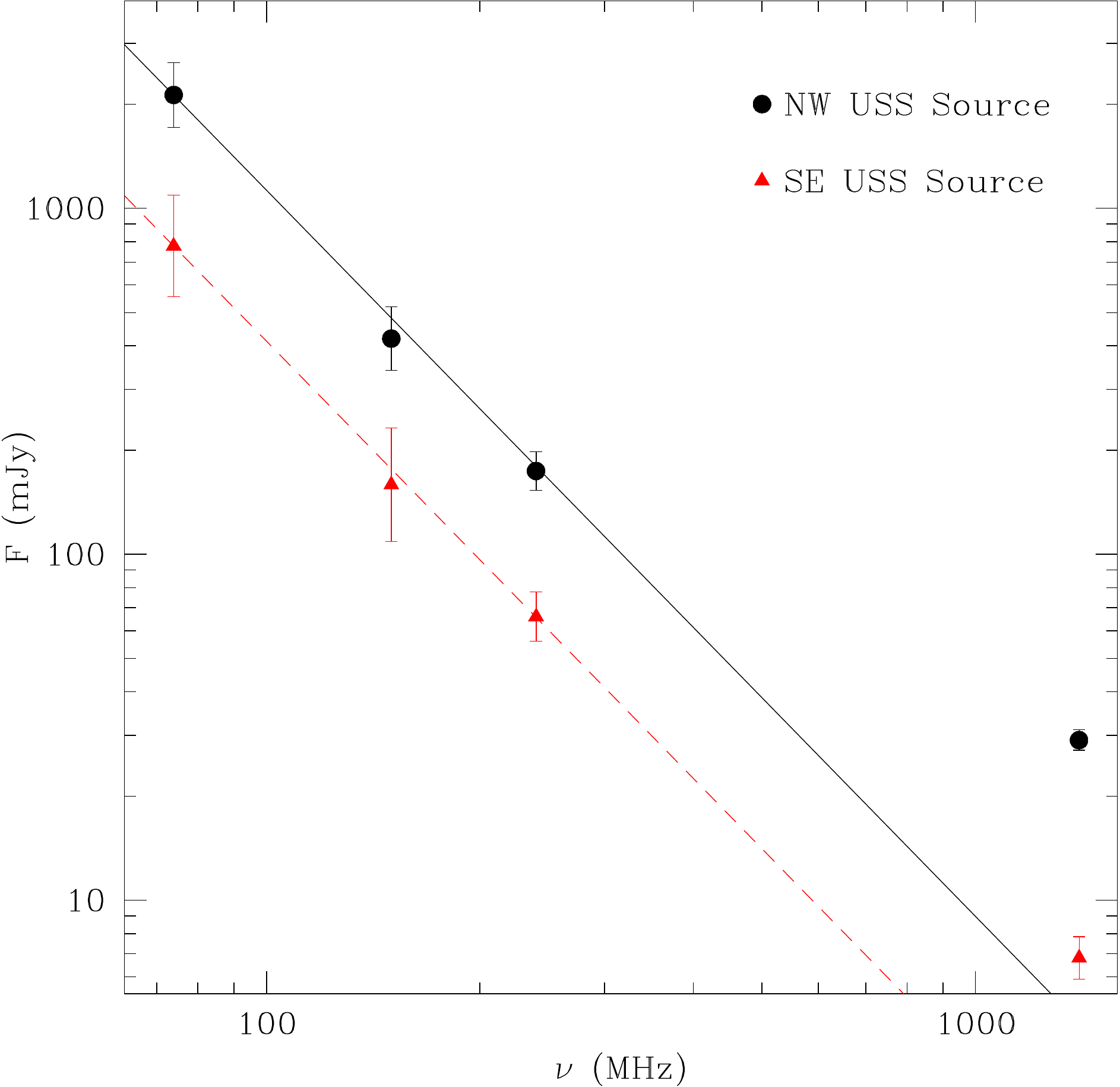}
\caption{
Radio spectral distribution in the region of the NW (black circles)
and SE (red triangles) USS radio sources (which are most likely radio
phoenixes).  Power-laws with $\alpha = -2.1$ (normalized at 74~MHz) are
overlaid.  The excess emission at 1.4~GHz is from the larger scale
radio relic, and is likely unrelated to the radio phoenixes, as
suggested by the uncorrelated morphologies (Figure~\ref{fig:radio}).
\label{fig:radio_spec}
}
\end{figure}

The radio spectra in the regions of the two USS sources are shown in
Figure~\ref{fig:radio_spec}, with lines of $\alpha = -2.1$ normalized at
74~MHz overlaid.  The spectra are consistent with power-laws at low
frequency, but lie significantly above the extrapolated models at
1.4~GHz.  This is consistent with the interpretation of the USS sources
as unrelated to the radio relic, which is most clearly visible at
1.4~GHz, as suggested by the very different morphologies at 74~MHz
and 1.4~GHz.  Assuming a typical spectral index of $\alpha = -1$ for
the radio relic, we find that the observations are consistent with a
non-detection at 74~MHz in the VLSSr.

Such diffuse, ultra-steep spectrum radio sources have been
identified in other 
clusters
\citep[e.g.,][]{2001AJ....122.1172S,2009A&A...508...75V,2010ApJ...722..825R,2011MNRAS.414.1175O,2011A&A...527A.114V,2012ApJ...744...46K,2013ApJ...772...84C},
but they are relatively rare, and the nature of these sources is not
fully understood.  Although we are unable to fit detailed models to
the radio spectra of these sources since we only detect them at three
frequencies, their steep spectra are consistent with old non-thermal electron
populations that have been re-energized by a shock (i.e., radio
phoenixes).  According to this model, a pre-existing electron population
(from, e.g., an old radio radio lobe) is re-energized by the passage of
an ICM shock.  Due to the high sound speed in the lobe, the plasma is
not shocked, but rather compressed adiabatically \citep{2001A&A...366...26E}.
This model can explain the lack of correlation between the low
and high frequency radio emission, and the fact that the USS sources
are displaced from the merger axis.  The morphology of these sources
depends on the spatial distribution of the pre-existing electron population, as
opposed to classical radio relics which directly trace ICM shocks.
The lack of obvious host galaxies for these sources is also consistent
with this interpretation, since older radio lobes would have had time
to detach from their hosts and fade before being re-energized by a
merger shock.

\subsubsection{Discussion} \label{sec:radio_discuss}

We conclude that the most likely interpretation of the observations is
that the diffuse radio emission is from at least two distinct types of
components: classical radio relics, where electrons are accelerated
by a shock through, for example, diffusive shock acceleration
\citep{1983RPPh...46..973D,2001RPPh...64..429M}, and radio phoenixes,
where old radio structures are re-energized by passing shocks through
adiabatic compression.  Both processes are driven by the same merger event.
Thus, CIZA~0107 provides a relatively rare case of radio relics and
radio phoenixes observed in the same system.
USS sources have been identified as radio phoenixes in other merging
systems \citep[e.g.,][]{2013ApJ...772...84C,2015MNRAS.448.2197D}, and
therefore may
provide a means of identifying dynamically disturbed systems from low
frequency radio observations.

It is currently unclear whether relics and phoenixes represent
different phases of a general ``life-cycle'' of non-thermal particles in the
ICM, or whether they arise from distinct particle populations.
I.e., radio lobes and phoenixes may eventually break apart, releasing
their particles to the ICM and providing a diffuse non-thermal
particle component that may be re-accelerated by ICM shocks to create
radio relics.  Alternatively, radio relic particles may have a
different origin, e.g., they may be accelerated directly from the
thermal pool.  However, it seems unlikely that diffusive shock
acceleration is efficient enough to accelerate particles from the
thermal pool to the required energies \citep{1998A&A...332..395E,2013MNRAS.435.1061P}.  Furthermore, not all strong
merger shocks are associated with radio relics
\citep{2011MNRAS.417L...1R}, suggesting that the presence of a preexisting
non-thermal particle population is required.
High resolution, low frequency radio observations will help confirm
the USS sources as radio phoenixes, which often have complicated,
filamentary morphologies \citep[e.g., see][]{2001AJ....122.1172S}.

\subsection{Inverse Compton Emission} \label{sec:ic}

In principle, the high temperature peak to the SW seen in
Figures~\ref{fig:tmaps}~\&~\ref{fig:ktprof} could arise from the
contribution of a non-thermal component that is not included in our
spectral model.  In particular, we expect some level of 
inverse Compton (IC) emission due to the interaction of the
synchrotron radio emitting 
electrons in the radio relic with the CMB.  Despite this expectation,
diffuse IC emission has yet to be conclusively detected in galaxy clusters
\citep[for recent results and reviews
see][]{2012RAA....12..973O,2014ApJ...792...48W,2015ApJ...800..139G}.
To test for the presence of IC emission, we added a power-law
component to the model for the high temperature bin roughly 330~kpc to
the SW shown in Figure~\ref{fig:ktprof}.  We tried both allowing the
temperature of the 
thermal component to vary and fixing it at the typical value at this
radius of 8~keV.  In neither case did including a power-law
component significantly improve the fit.

To place a conservative limit on IC emission from this region, we fit
the spectrum with an absorbed power-law, with no thermal component.
This model provided a statistically equivalent fit to the single
temperature thermal model, with an F-test probability of 72\% (we consider
an F-test probability of $\la 5$\% to indicate a significant
improvement).
This degeneracy is due to our fairly shallow exposure, and to the lack
of strong emission lines from such high temperature thermal plasma.
The best-fit photon index was $\Gamma = -1.5^{+0.1}_{-0.1}$, which is
close to the spectral slope of the radio emission $\alpha^{240}_{74} =
-1.3$ (Section~\ref{sec:radio}).  Assuming that the synchrotron and IC
emitting particles are the same population, and that these particles follow a
power-law distribution in energy, the synchrotron and IC spectral
slopes are expected to be equal \citep{1977OISNP..89.....P}.  This
model gives a total 2--10~keV flux of $2.4^{+0.1}_{-0.1} \times
10^{-12}$~erg~cm$^{-2}$~s$^{-1}$.

\citet{2001ApJ...557..560P} gives a convenient expression relating the
monochromatic IC X-ray and synchrotron radio fluxes to the implied
magnetic field strength, 
\begin{equation} \label{eq:flux_ratio}
\begin{aligned}
R \equiv \frac{f_{\rm IC}(kT)}{f_{\rm sync}(\nu)} & = 1.86 \times 10^{-8}
  \left( \frac{\rm photons}{\rm cm^{2}\, s\, keV\, Jy} \right) \\ &
  \times \left (\frac{kT}{20\, {\rm
      keV}} \right)^{-\Gamma}\left( \frac{\nu}{\rm{GHz}} \right)^{\Gamma - 1} \\ &
  \times \left( \frac{T_{\rm CMB}}{2.8 {\rm K}} \right)^{\Gamma + 2} \left(\frac{B}{\mu
    {\rm G}} \right)^{-\Gamma}c(p),
\end{aligned}
\end{equation}
where $\Gamma = (p + 1)/2$,  $p$ is the power-law slope of the electron energy
distribution $N(E) \propto E^{-p}$, $f_{\rm IC}(kT)$ is the IC flux
density at energy $kT$,
$f_{\rm sync}(\nu)$ is the synchrotron flux density at
frequency $\nu$, $T_{\rm CMB}$ is the CMB temperature at the
cluster redshift, and $c(p)$ is a normalization factor that is a
complicated function of $p$ \citep[with values $10 < c(p) < 1000$ for typical
values of $p$, see][]{1979rpa..book.....R}.  
For $2 \la p \la 5$, $c(p)$
can be approximated as $c(p) \approx e^{1.42 p - 0.51}$.  Using this
approximation with Equation~\ref{eq:flux_ratio} and solving for the
magnetic field, one finds
\begin{equation} \label{eq:bfield}
\begin{aligned}
&B = \left( \frac{20 {\rm keV}}{kT} \right) \left( \frac{\nu}{{\rm
      GHz}} \right)^{(p-1)/(p+1)} e^{\frac{2.84(p-r)}{p+1}} \mu \rm{G}, \\ &
r = 0.7 \ln \left[ \frac{R_{\rm obs}(kT, \nu)}{1.11 \times 10^{-8}}
\right] .
\end{aligned}
\end{equation}
Note that there is a typo in the exponential term in the equivalent
expression in \citet{2001ApJ...557..560P} (their Equation~7).
Our power-law fits in this region give an IC flux density of 
$F_{\rm IC}(1 \, {\rm keV}) = 4.8 \times 10^{-4}$
photons~cm$^{-2}$~s$^{-1}$~keV$^{-1}$, while the radio observations
give $F_{\rm sync}(1.4 \, {\rm GHz}) = 1.35 \times 10^{-2}$~Jy (for
the latter, bright point sources have been removed to give the flux of
the diffuse emission only).
For $\Gamma=1.55$, this gives a magnetic field strength of $B =
0.01$~$\mu{\rm G}$.

Since the X-ray surface brightness does not correlate with the radio
emission in detail, and since diffuse IC flux has yet to be
conclusively detected
in deep observations of other galaxy clusters, we conclude that the
thermal model is much more likely, despite the fact that the thermal
and non-thermal models provide statistically equivalent fits.
Since the non-thermal model assumes that all
of the emission in this region is IC emission, whereas it is in fact
very likely dominated by thermal ICM emission, the above flux is a very
conservative upper limit on the true IC flux, and thus $B >
0.01$~$\mu{\rm G}$ is a very conservative lower limit on the magnetic
field strength. 
For comparison, the equipartition magnetic field
  strength implied by the total diffuse 1.4~GHz flux is roughly
  1.6~$\mu{\rm G}$, assuming a ratio of energy in protons to energy in
  electrons of $k=1$ and a minimum Lorentz factor of $\gamma_{\rm min}
  = 100.$


\section{Summary}

We present results from X-ray, optical and radio observations of the
massive galaxy cluster CIZA~J0107.7+5408.  Observations at all three
wavelengths show a double-peaked morphology, with all peaks lying
along roughly the same axis.  The optical and 1.4~GHz radio peaks are at
larger cluster radii than the X-ray peaks.  The X-ray temperature map reveals
a high temperature peak to the SW, roughly coincident with the SW
radio peak at 1.4~GHz.  We conclude that this system is a 
post core passage dissociative merger.  The X-ray peaks lag the
optical galaxy density peaks due
to ram pressure forces on the ICM.  Merger shocks lead the merging
subclusters, giving rise to the possible doubly peaked
radio relic and the
shock heated gas in the region of the SW relic.  The SW temperature
rise implies a shock Mach number of at least $M = 2.3 \pm 0.4$.
Rough estimates suggest that follow up optical lensing and
spectroscopic observations may allow interesting limits to be placed
on the self-interaction cross-section of dark matter, as has been done
for other dissociative merging systems.

Low frequency radio observations reveal diffuse, ultra-steep spectrum
radio emission, with $\alpha \approx -2$.  This emission shows two
peaks near the SW subcluster,
although the peaks do not correlate with the merger axis, the X-ray
emission, or the high frequency radio emission (\ie, the radio
relics).  We suggest that these features are radio phoenixes, formed
when old but relatively cohesive radio structures (likely created by
radio galaxies) are re-energized due to adiabatic compression by
passing merger shocks.  Thus, CIZA~0107 is a relatively rare case
containing clear examples of both classical radio relics and USS radio
phoenixes.

Finally, we use the X-ray observations to place very conservative
upper limits on the IC flux, and lower limits on the ICM magnetic
field strength, in the region of the SW relic.

\section*{Acknowledgments}

Support for this work was partially provided by the Chandra X-ray Center
through NASA contract NAS8-03060, the Smithsonian Institution, and by
the {\it Chandra} X-ray Observatory grant GO3-14134X.
Basic research in radio astronomy at the Naval Research Laboratory is
supported by 6.1 Base funding.
This research has made use of the NASA/IPAC Extragalactic Database
(NED), which is operated by the Jet Propulsion Laboratory, California
Institute of Technology, under contract with the National Aeronautics
and Space Administration.
We thank the staff of the GMRT who have made these GMRT observations
possible. GMRT is run by the National Centre for Radio Astrophysics
of the Tata Institute of Fundamental Research.
The National Radio Astronomy Observatory is a facility of the National Science
Foundation operated under cooperative agreement by Associated Universities,
Inc.
We thank Dale Kocevski for providing spectroscopic redshifts for the
BCGs, and Paul Nulsen for useful discussions.

\bibliographystyle{apj}
\bibliography{references}
\end{document}